\documentclass[10pt, nocopyrightspace]{sig-alternate-10pt-mobicom}

 \pdfoutput=1
\include{psfig}  
\usepackage{epsfig}
\usepackage{algorithm }
\usepackage{xspace}
\usepackage{amssymb}
\usepackage{amsmath}
\usepackage{algpseudocode}
\usepackage{float}
\usepackage{subcaption}
\usepackage{graphicx}
\usepackage[update,prepend]{epstopdf}
\usepackage{flushend}
\usepackage{color}
\usepackage{url}
\usepackage{array}

\floatstyle{plain}
\newfloat{deneme}{t!}

\newcommand \be {\begin{equation}}
\newcommand \ee {\end{equation}}

\algnewcommand{\LineComment}[1]{\State \(\triangleright\) #1}

\newcommand{\system}{\textsc{SkyLiTE}\xspace}
\newcommand{\sysran}{\textsc{SkyRAN}\xspace}
\newcommand{\syshaul}{\textsc{SkyHaul}\xspace}
\newcommand{\syscore}{\textsc{SkyCore}\xspace}

\begin{document}

\hyphenation{thro-ugh-put o-f-d-m-a plan-ned program-mable draf-ted wi-max two-di-men-sion-al para-digm}

\title{SkyLiTE: End-to-End Design of Low-altitude UAV Networks for Providing LTE Connectivity}

\numberofauthors{1}
\author{
\alignauthor Karthikeyan Sundaresan, Eugene Chai, Ayon Chakraborty, Sampath Rangarajan\\
       \affaddr{NEC Labs America, Princeton, NJ 08540}\\
       \email{\large{Email: karthiks@nec-labs.com}}
	}

\maketitle 
\vspace*{0.25cm}

\begin{abstract}
Un-manned aerial vehicle (UAVs) have the potential to change the landscape of wide-area wireless connectivity by bringing them
to areas where connectivity was sparing or non-existent (e.g. rural areas) or has been compromised due to disasters.
While Google's Project Loon and Facebook's Project Aquila are examples of high-altitude, long-endurance UAV-based connectivity efforts in this direction, the telecom operators (e.g. AT\&T and Verizon) have been exploring low-altitude UAV-based LTE solutions for on-demand deployments.
Understandably, these projects are in their early stages and face formidable challenges in their realization and deployment. 
The goal of this document is to expose the reader to both the challenges as well as the potential offered by these unconventional connectivity solutions. We aim to explore the end-to-end design of such UAV-based connectivity networks particularly in the context of low-altitude UAV networks providing LTE connectivity. Specifically, we aim to highlight the challenges that span across multiple layers (access, core network, backhaul) in an inter-twined manner as well as the richness and complexity of the design space itself. To help interested readers 
navigate this complex design space towards a solution, we also articulate the overview of one such end-to-end design, namely
 \system -- a self-organizing network of low-altitude UAVs that provide optimized LTE connectivity in a desired region. 

\end{abstract}


\section{Vision}
\label{VISION}
\begin{figure}
\hspace*{0.25cm}
    \centering
    \includegraphics[width=\columnwidth]{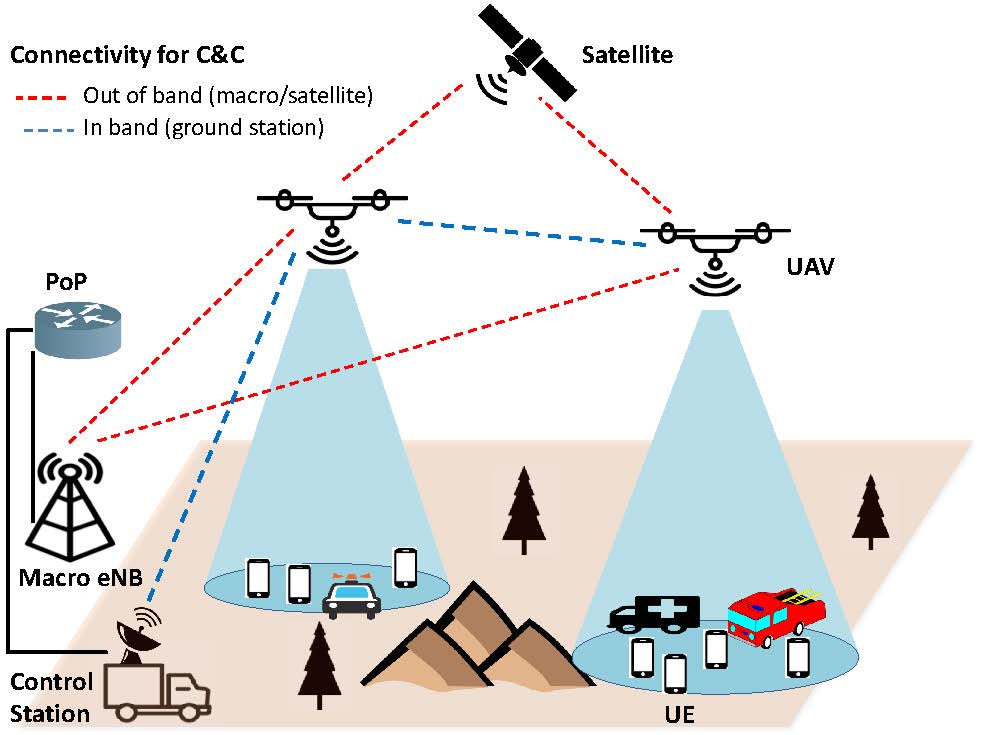}
    \caption{Low-altitude platform/UAV Network.\label{fig:lap}}
\end{figure}

Today, wireless access and connectivity is largely a two-dimensional (terrestrial) problem,
where  well-planned base stations are statically deployed in economically viable areas. The growing maturity of 
unmanned aerial vehicle (UAV) technology aims to change that notion by adding a third 
spatial degree of freedom (aerial), which has the potential to completely change the landscape
of wireless connectivity. We now have the technical means to deploy aerial base stations (BSs on UAVs)
on the fly and provide wireless connectivity in areas where a pre-existing connectivity infrastructure does not exist or exists sparingly (e.g. rural areas)
or  has been compromised (e.g. man-made or natural disasters). 
{\em Our overarching vision is to be able to realize such UAV networks that are capable of providing on-demand, wide-area (spanning one
or more cities) wireless connectivity using the most
popular wireless access technology today, namely LTE.} (Fig.~\ref{fig:lap}).

\section{Where Are We Today?}
\label{TODAY}

The tight regulation of the commercial airspace by federal authorities coupled with the scope and longevity of the connectivity solutions envisioned, has given rise
to two category of efforts.

\begin{figure}
    \begin{minipage}[t]{0.55\columnwidth}
        \centering
        \includegraphics[width=\columnwidth,height=4cm]{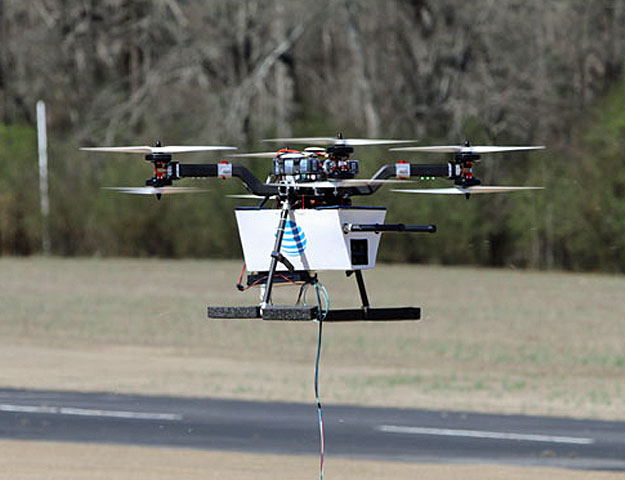}
        \caption{AT\&T's Tethered Cell on Wings.
        \label{fig:attcow1}}
    \end{minipage}
    \hspace{5pt}
    \begin{minipage}[c]{0.40\columnwidth}
    	\vspace*{-1cm}
        \centering
        \includegraphics[width=\columnwidth,height=2cm]{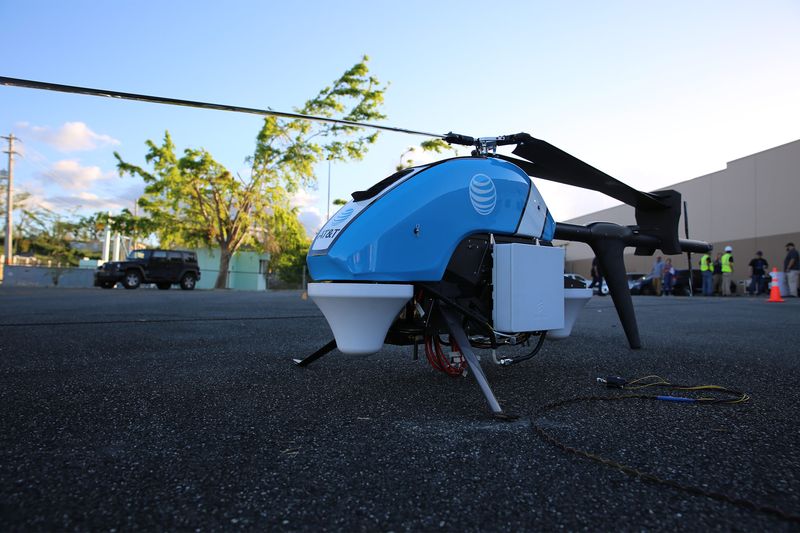}
        \caption{AT\&T's CoW in Peurto Rico.\label{fig:attcow2}}
        \vspace{2pt}
        \includegraphics[width=\columnwidth,height=2cm]{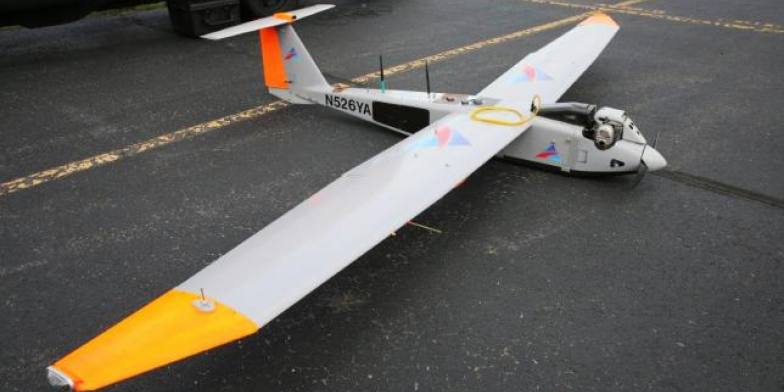}
        \caption{Verizon's Cell on Wings.\label{fig:vzwcow}}
    \end{minipage}

\end{figure}

\begin{itemize}
\item{High-altitude Platform (HAP) networks:}
These UAV networks aim to provide connectivity solutions to the un-connected parts of the world.
This requires providing connectivity over large geographical regions over longer periods of time, which can be accomplished by operating at high altitudes  (for wider coverage) in a cost and energy efficient (with light-weight, power-efficient UAVs) manner. Google's Project Loon~\cite{Loon} (employs balloons, Fig.~\ref{fig:loon}) and Facebook's Project Aquila~\cite{Aquila} (employs custom drone, Fig.~\ref{fig:aquila}) are efforts in this direction. They operate above the regulated airspace in the stratosphere at altitudes between 20-50 Km, and aim to leverage the existing stratified air currents in the atmosphere in different directions to move and position the UAVs appropriately with minimal energy and provide connectivity in desired regions. Recently, Project Loon marked an important milestone by showcasing its feasibility to provide LTE connectivity (from its balloons) in Puerto Rico in the aftermath of hurricane Maria. 

\begin{figure}
    \centering
    \includegraphics[width=0.8\columnwidth]{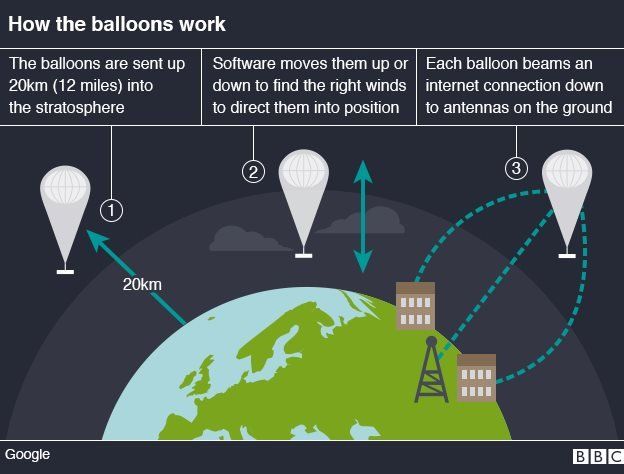}
    \caption{Google's Project Loon [Source: BBC, Google].\label{fig:loon}}
\end{figure}

\begin{figure}
    \centering
    \includegraphics[width=0.8\columnwidth]{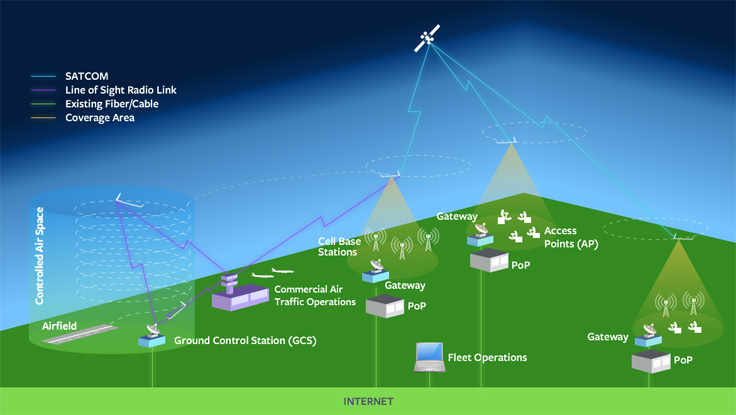}
    \caption{Facebook's Project Aquila [Source: Facebook].\label{fig:aquila}}
\end{figure}

\item{Low-altitude Platform (LAP) networks:}
These UAV networks aim to provide wireless connectivity solutions on-the-fly in regions where additional capacity/coverage is needed (e.g. stadiums, concerts, etc.) or existing connectivity infrastructure is overwhelmed or compromised (e.g. man-made or natural disasters). 
By targeting short-term connectivity solutions over limited geographic regions, these solutions operate in a much smaller scale at lower altitudes
(in the troposphere from several hundred meters to a few Km) compared to their high-altitude counterparts. 
These have generated significant interest from major telecom operators like AT\&T and Verizon, who have conducted their own trials~\cite{AttCow,VzwCow} (Figs.~\ref{fig:attcow1},~\ref{fig:vzwcow}) to understand the feasibility of  providing LTE connectivity from a low-altitude UAV, also termed as CoWs (Cell on Wings). 
Similar to Project Loon, recently, AT\&T deployed a CoW using a tethered (for power and data) helicopter at 200 ft to provide temporary LTE  service in Puerto Rico in the aftermath of  hurricane Maria~\cite{AttCowPR} (Fig.~\ref{fig:attcow2}). The European Union is investigating the feasibility of a LAP based LTE network that employs tethered balloons (HeliKite) in its {\small ABSOLUTE} project~\cite{EU-Absolute}.

\end{itemize}

Understandably, these efforts are in their initial stages. 
Indeed, we are yet to comprehensively understand how to optimize even a single UAV for providing LTE connectivity in practical deployments (be it for high-altitude or low-altitude), let alone manage a {\em network} of UAVs to offer seamless connectivity over a much larger area.
The goal of this document is to shed light on the various challenges that face the design of these UAV networks. 
We will discuss them in the context of LAP networks (employing rotary wing aircrafts) that provide LTE connectivity. 
Our discussions also apply when fixed wing aircrafts are employed as UAVs.
However, note that fixed wing aircrafts need to maintain continuous forward motion to remain aloft. 
Hence, compared to rotary wing aircrafts that can stay stationary, their constant mobility 
introduces an additional layer of challenges and design considerations arising from UAV path planning that we discuss in Section~\ref{FIXED}. 
Further, while several of the design challenges and elements discussed for LAP networks  will equally apply to HAP networks, there are some significant differences as well between the two that necessitate weighing certain tradeoffs and accompanying decisions differently. We will discuss these in Section~\ref{HAP}.



%

\section{Architecture}
\label{ARCH}
The network architecture of an untethered,  low-altitude UAV network is shown in Fig.~\ref{fig:lap}.  
Note that one could increase the operational lifetime of the network by having the UAVs tethered to a power/data source and carrying only essential radio equipment and/or antennas in the air. However, being tethered to a ground vehicle would significantly restrict its deployment flexibility to only accessible areas on the ground (potentially less useful in disasters), as well as its coverage and optimization capabilities. Hence, we will focus on the untethered scenario to explore the most flexible version of UAV network deployments.

The UAV carries the LTE base station (eNB) and provides connectivity to users (UEs) on the ground. 
There are two potential realizations of our network vision that can cater to different use cases.  
(i) {\em Stand-alone}: We can deploy a stand-alone LTE connectivity infrastructure, which does not connect (backhaul) to the Internet (i.e. an Internet point-of-presence does not exist).
 In cases, where no pre-existing connectivity infrastructure is available, such a stand-alone network can be useful in providing
 connectivity and communication between first responders, emergency services and people in disaster scenarios. 
 (ii) {\em Internet-backhauled}: When a point-of-access (such as a ground station or macro-cell) to Internet exists, then in addition to the stand-alone services that can be delivered, one can also provide Internet access to both users in emergency scenarios as well as those in areas of limited/no connectivity.
 
 There are two components to wireless connectivity in this architecture:  the wireless links between the base station on the UAV and the UEs on the ground constitute the radio {\em access} network (RAN), while the wireless links between the UAVs themselves that form a wireless multi-hop mesh network  (in the air) constitute the {\em backhaul}. While both these components cater to {\em data} communication between users and Internet, there is also another wireless connectivity component that is needed to provide {\em command and control} (C\&C) for the UAVs. 
 One could potentially leverage the RAN and backhaul connectivity for C\&C as well (in addition to data/payload). However,  given the mobile and multi-hop nature of the backhaul, the latter may not ``always" guarantee connectivity for safety-critical operations such as UAV C\&C (crash avoidance, .  
 Hence, the preference for reliable C\&C connectivity to the UAVs is typically through a static, always-reachable network node. This can be achieved through macro cell towers (less expensive option), when one or more is available and can reach all the UAVs, or through a satellite (more expensive option) as shown in Fig.~\ref{fig:lap}. The latter, being the only viable option for C\&C in high-altitude networks, is adopted by Project Loon and Aquila.

\section{Layered Challenges}
\label{CHAL}
The goal of this section is to explore the design space of our network vision by understanding the various layered challenges that arise in its realization.  
A conventional LTE network consists of two main components: static base stations that provide wireless access to users (RAN - radio access network),
and a wired core network of gateways (EPC - evolved packet core) that sits behind the base stations and is responsible for all the mobility, management and control functions, as
well as routing user traffic to/fro between users and the Internet.
To begin with, deploying and managing an LTE network (RAN + EPC) is no ordinary feat.
Couple this with the notion of deploying a LTE network on aerial base stations, namely UAVs, which are highly restrictive in their compute capabilities, endurance, operational time, and payload capacity -- we have an additional layer of inter-twined 
challenges. To complicate things further, the connectivity for the core network now becomes wireless (compared to wired before), which makes the design of the multi-hop wireless backhaul for the core network challenging in its own right. 
We now discuss these challenges in the context of RAN, backhaul and core design for UAV networks. 

\begin{figure}
    \centering
    \includegraphics[width=\columnwidth]{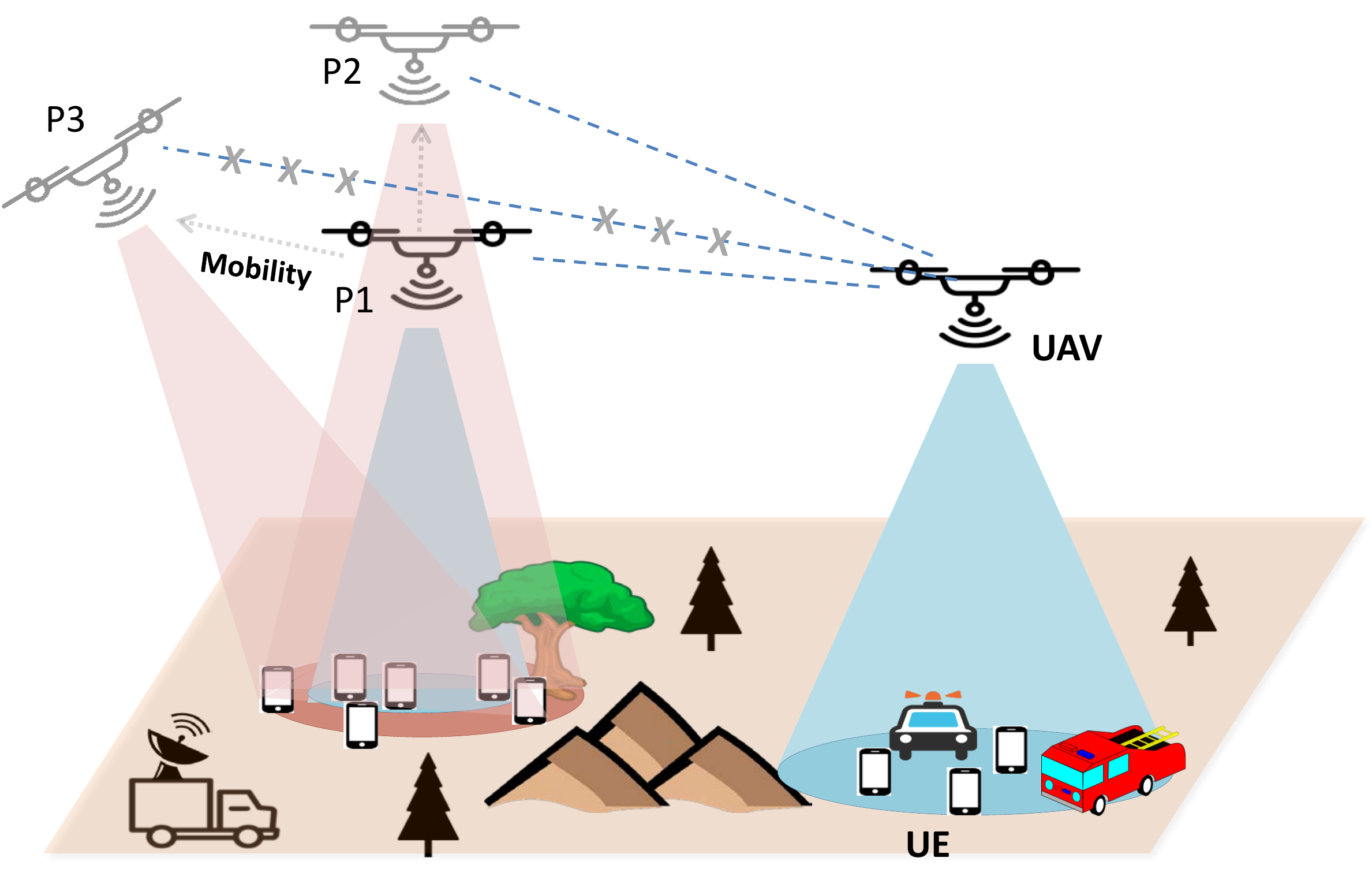}
    \caption{Coupling between RAN and Backhaul: Going from P1 to P2 increases RAN coverage, but some UEs are still affected by shadowing. Going to P3 provides best coverage but affects backhaul connectivity.\label{fig:ranhaul}}
\end{figure}

%

\subsection{RAN Challenges}
The key requirement for our UAV-driven RAN is to position the UAVs in 3D space so as to create multiple LTE cells on-demand and provide adequate coverage (e.g. above a certain minimum rate) and capacity to all devices/users in the region of interest (Fig.~\ref{fig:multiuav}). Given the shorter altitudes of our network (less than 400 ft),
the wireless link between the UAV and UEs on the ground is subject to various obstacles (trees, foliage, buildings, houses, etc.), which can affect the signal strength and data rates on the link significantly across various positions of the UAV as well as across different UEs~\cite{uav-link1,uav-link2}.  
Hence, realizing an optimized network of UAVs on-demand, requires one to first construct an RF map that characterizes the RF channel conditions between the UAVs and the devices in the entire environment of interest.
We now discuss the various challenges from the perspective of a single UAV trying to optimize its own cell, followed by the inter-dependent
challenges across UAVs.

   \begin{figure}
        \centering
        \includegraphics[width=\columnwidth]{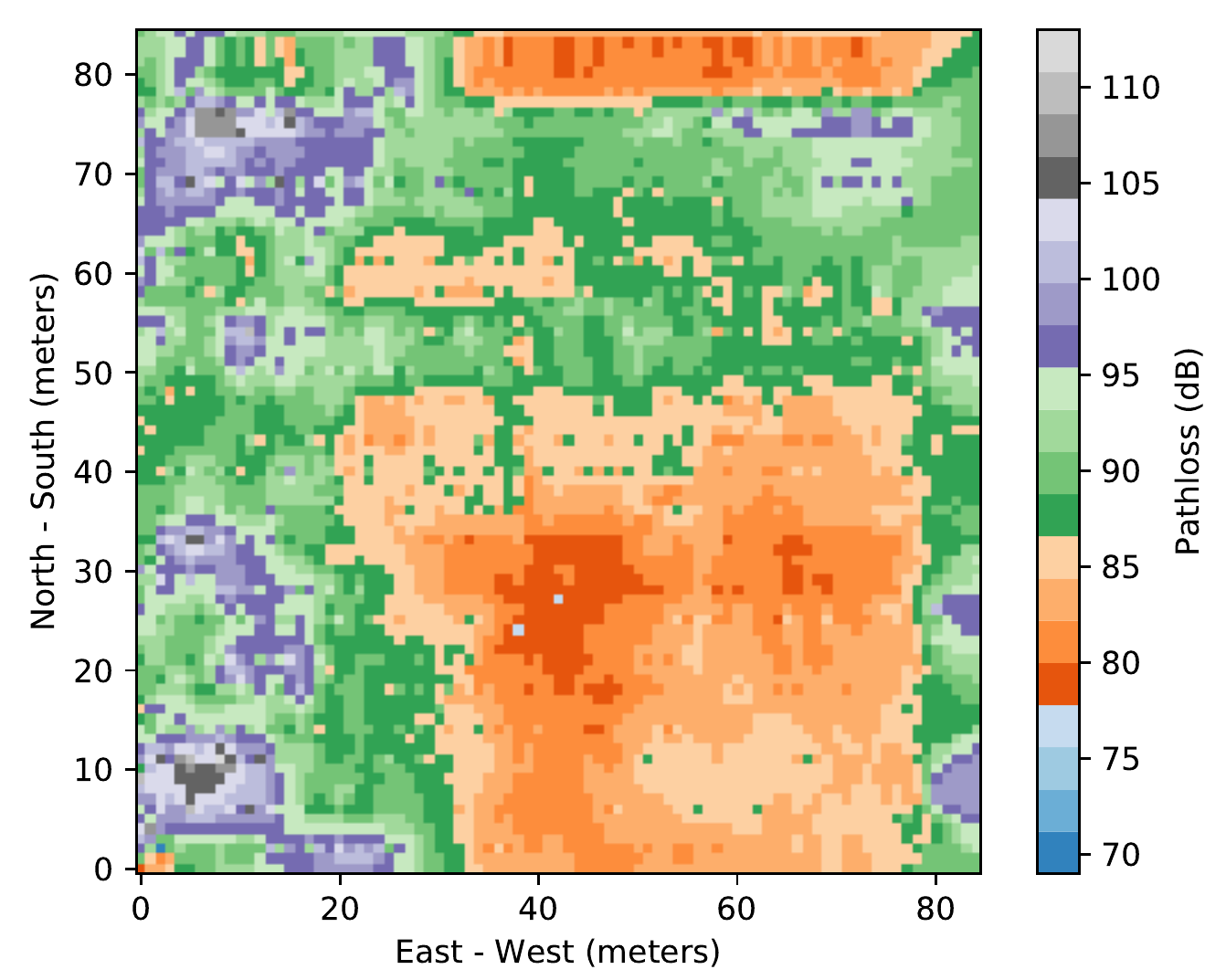}
        \caption{RF map for one UE at a given UAV altitude.
        \label{fig:rfmap}}
    \end{figure}

\subsubsection{Localizing Devices}
Constructing an RF map to determine an appropriate position for the UAV, requires the UAV to conduct signal strength (channel) measurements to devices. To accomplish the latter, it is essential for the UAV to first estimate the location of the devices to whom the measurements are being made. 
In addition, such location data provides valuable information to first responders in locating and assisting users directly in emergency scenarios. 
Several devices today have GPS functionality, but RANs do not have a direct API yet  
to collect and use the GPS information of its devices in its connectivity decisions. While an obvious fix is to obtain such location information through OTT applications, 
the challenge is to also cover devices that may not have GPS or have not enabled their location services.
In essence, {\em Can we leverage just the LTE RAN to automatically localize its devices without relying on their GPS functionality?}

\begin{figure}
\centering 
\begin{subfigure}{0.47\columnwidth}
 \centering
        \includegraphics[width=\columnwidth,height=4cm]{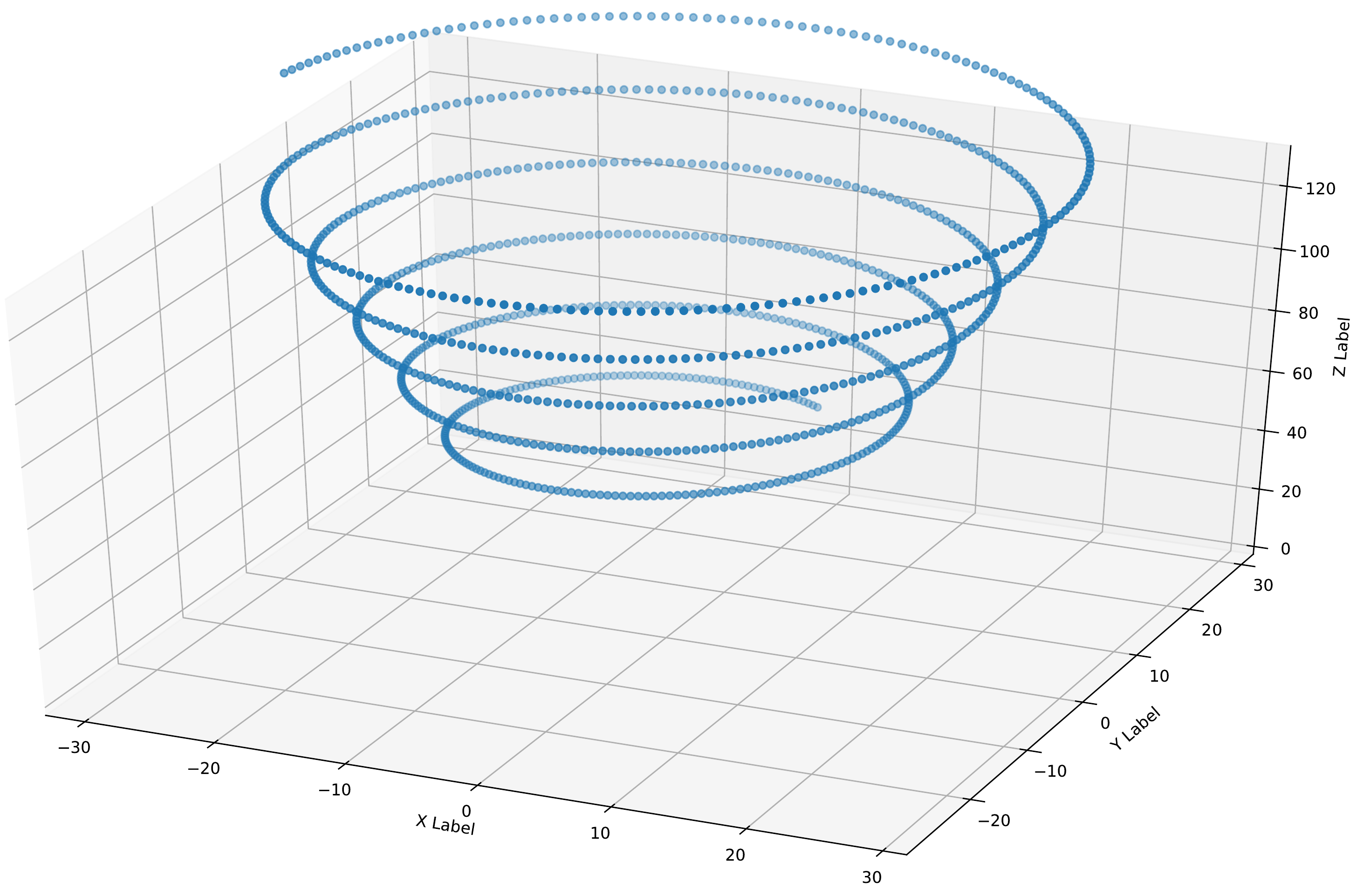}
\end{subfigure}
\begin{subfigure}{0.47\columnwidth}
 \centering
        \includegraphics[width=\columnwidth,height=3cm]{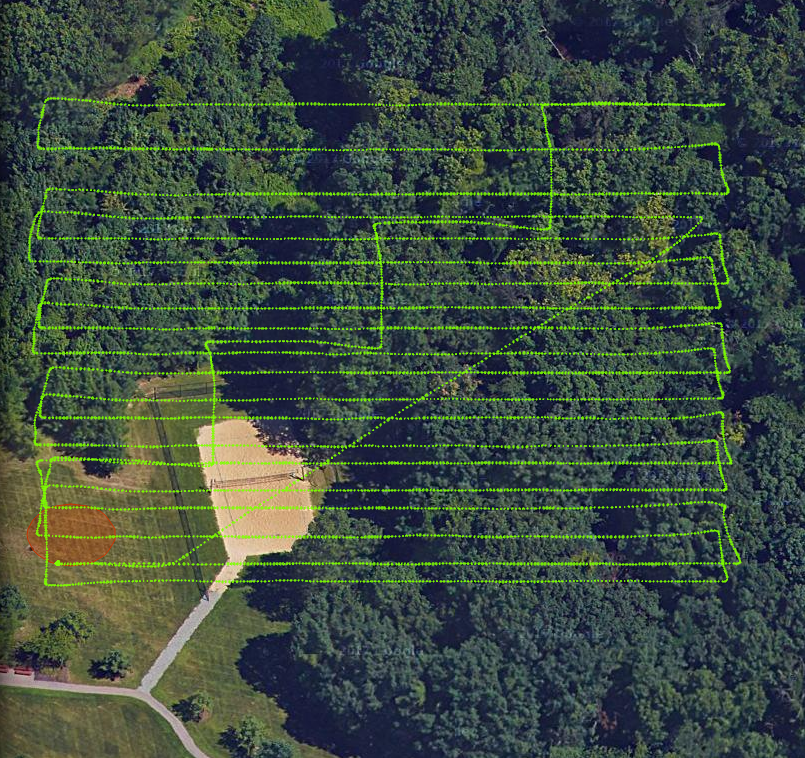}
\end{subfigure}
\caption{Examples of airspace sampling for RF map construction.\label{fig:airspace}}
\end{figure}

\subsubsection{RF Map Construction}
In traditional RAN deployments, BSs are deployed after detailed offline site survey/planning and RF measurements. 
In contrast, UAV networks are expected to be ``on-demand" mobile deployments. 
An RF map (similar to a heat map) is specific to each client (UE) and captures the single strength to the UE on ground from different positions of the UAV in air. 
To construct an accurate RF map (as shown in Fig.~\ref{fig:rfmap}), one would need the UAV to first identify the operational airspace (Fig.~\ref{fig:airspace}) from which it can provide coverage to a desired area on the ground; then move around this entire (identified) airspace and conduct channel measurements to all its devices in this area (whose locations have already been estimated) from each of those positions. 

Realizing both these steps is challenging for the following reasons. Note that the coverage provided by a UAV on the ground, depends on how the radiation pattern of its antenna (mounted on the UAV) illuminates the ground. Hence, as the UAV moves, the coverage provided by its antenna on the ground will also change. Thus, only from certain positions in the airspace with appropriate orientations of its antenna (yaw/tilt), will the UAV be able {\em to provide sustained coverage for a desired area on the ground, making it challenging to identify the operational airspace for the UAV}.
Further, conducting measurements over the entire operational airspace, which could be potentially large, would incur significant overhead (energy drain for the UAV) and latency. The challenge here is {\em to construct a reasonably accurate RF map of the environment with only a limited number (sampled from the operational airspace) of RF measurements to the devices}.

\subsubsection{UAV Positioning and Orientation}
With the help of the RF map generated, the UAV would then need to solve a challenging optimization algorithm to determine its appropriate position and orientation that best serves a multitude of devices simultaneously.  Consider a simple
example as shown in Fig.~\ref{fig:ranhaul}. As the UAV changes its altitude (from P1 to P2) and/or its antenna orientation (from P1 to P3) through its yaw/tilt, this changes the coverage pattern on the ground. The closer the UAV 
is to the ground, the smaller the set of devices it covers, and vice versa. On the other hand, the closer the UAV is to the ground, the greater
the signal strength received at its devices and hence better data rates to the devices it covers. Thus, while being closer to the ground is beneficial for
certain clients that are covered by the antenna pattern, it hurts performance for certain other clients who cannot be effectively covered.
Hence, the challenge for the optimization problem is in {\em catering to multiple devices simultaneously}.

\begin{figure}
    \centering
    \includegraphics[width=\columnwidth]{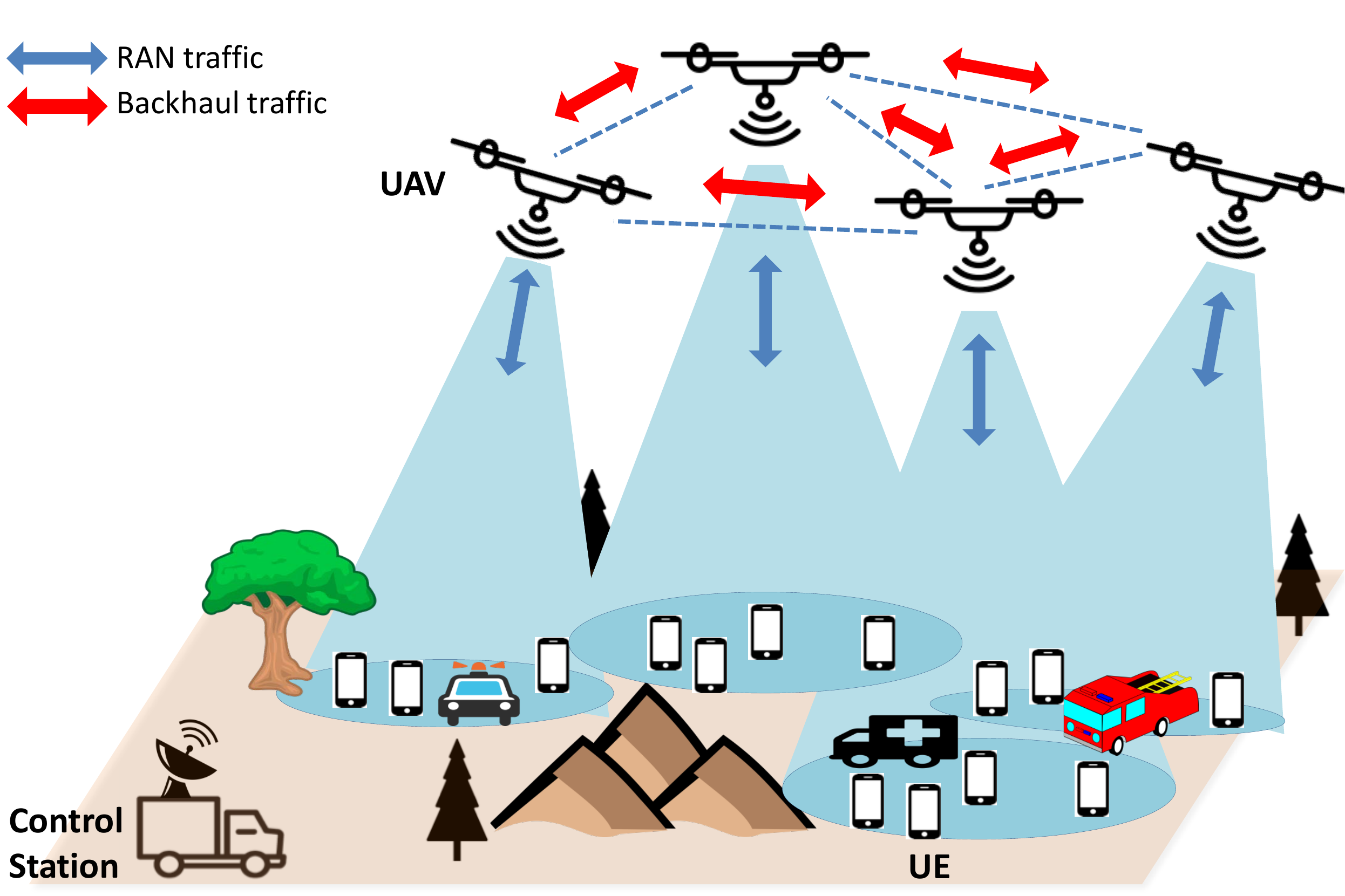}
    \caption{Coordinated Multi-UAV Network.\label{fig:multiuav}}
\end{figure}

\subsubsection{Multi-UAV Coordination}
So far  we discussed about how to provide coverage to clients from a single UAV. 
Often, a single UAV may not be sufficient when UEs are spread over a large region, requiring the use of multiple UAVs as shown in Fig.~\ref{fig:multiuav}.  
With the introduction of multiple UAVs, the problem now becomes significantly more challenging. 
Several questions arise pertaining to {\em how these multiple UAVs should be ``jointly" deployed to efficiently cover the entire area of interest}. 
Note that the position and resulting coverage provided by one UAV affects the choice of the position/coverage of other UAVs. 
While it is optimal to centrally solve for the position/coverage of all UAVs {\em jointly}, this is hardly feasible in practice (in real-time), given the coordination required and the challenge of optimizing even the position of a single UAV (as discussed above). Hence, a more distributed and scalable approach to self-organization of the UAVs is needed. On the other hand, such distributed approaches, need to ensure that as the UAVs go about conducting their RF measurements
and providing connectivity, their trajectories do not collide with one another. Further, in addition to providing RAN coverage, the UAVs also need
to wirelessly mesh with one another to establish a robust, multi-hop backhaul. Hence, a distributed optimization of UAVs for RAN coverage may
not be in the best interest of their backhaul optimization and vice versa (we discuss this in more detail in Section~\ref{sec:backhaul}).
Finally, wide-area networks need to deal with mobility  of the UEs, which result in expensive (latency, control overhead) handoff events (moving from one cell to another). Now, with the base stations (UAVs) themselves being
mobile, we need to ensure that the changing coverage of the UAVs (due to their mobility) does not trigger unnecessary handoff events, especially for static UEs.

\begin{table*}[t]
\centering
\caption{Comparison between backhaul connectivity modalities.}
\resizebox{\textwidth}{!}{

\begin{tabular}{l|c|c|c|c|c|c}
Challenge & \textsc{Sub-6 LTE} & \textsc{Sub-6 WiFi} & \textsc{mmWave-60 GHz}  & \textsc{mmWave-28/39/70/80 GHz} & \textsc{FSO} \\
\hline
\textbf{Cost} & Low & Low & Low & High & Very High  \\
\textbf{Energy Consumption} & Low & Low & Low & High & Very High \\
\textbf{Form Factor} & Small & Small & Small & Large & Large \\
\textbf{Interference} & High & High & Low & Low & Very Low \\
\textbf{Communication Range} & Low-Moderate & Low-Moderate & Low & High & Very High \\
\textbf{Robustness of Topology/Routing} & High & High & Low & Low & Very Low \\
\textbf{Spectrum Management Overhead} & High & High & Low & Low & Low \\
\textbf{Spectrum Licensing} & Licensed/Unlicensed & Unlicensed & Unlicensed & Licensed/Unlicensed & No Licensing

\end{tabular}
}
\label{table:compbackhaul}
\end{table*}


\subsection{Backhaul Challenges}
\label{sec:backhaul}
Recall that while the RAN is responsible for delivering/receiving traffic directly from the UEs, the backhaul (connectivity between UAVs) is 
responsible for getting this traffic to/from the Internet or other UEs in the network. The key requirement for the backhaul is to organize the UAVs in the air such that they can form a high capacity wireless mesh backhaul that can carry all the traffic demand imposed by the UEs to/from the RAN as shown in Fig.~\ref{fig:multiuav}. This would include not just the positioning of the UAVs, but being a wireless, multi-hop backhaul, other UAV configuration parameters relating to connectivity and wireless interference, such as antenna orientation, spectrum assignment, wireless technology for operation, etc. need to be factored in as well. Being an equally important part of the connectivity fabric of
UAV networks, it is critical to understand the challenges underlying the realization and deployment of such a multi-hop wireless mesh backhaul in the air. 

\subsubsection{Coupled vs. De-coupled Design}
In a traditional LTE network, only the access (RAN) is wireless, while the connectivity from the base stations to the core network (which connects to Internet) is a high-speed, reliable wired network. 
However, in a UAV network, traffic encounters two wireless components (access and backhaul) before it can reach the Internet or other UEs. 
With the UAVs being the common nodes that anchor both the RAN and backhaul, the RAN and backhaul performance are inherently coupled
and together determine the capacity of the UAV network.
Given the challenges in UAV deployment even from an isolated RAN perspective, it might be tempting to consider a backhaul 
design that is decoupled from that of the RAN. However, such an approach can unfavorably affect the backhaul and consequently the end-end performance of the network as a whole. Hence a coupled design is definitely in the best interest of the whole network. 

 A coupled design, however, is not without its fair share of problems. Note that the RAN deals with individual wireless links to UEs on the ground.
 Hence, its channel dynamics (multi-path fading, shadowing, UE mobility, etc.) change at a much finer time scale (millisecs) compared to that of the backhaul wireless links (secs to minutes) that carry aggregate traffic (of multiple cells) between UAVs in the air. Hence, realizing a joint optimization of the RAN and backhaul performance to compute the optimal UAV network
 configuration, requires obtaining relevant (channel) information from all access and backhaul links at a central location. For the computed
 configurations to be relevant for observed conditions, it is necessary to realize and execute this joint optimization at the granularity of milliseconds, which is
 practically infeasible. Further, with the environment being subject to UAV dynamics (UAVs going offline for energy replenishment),
 such optimization has to be invoked frequently, posing an overhead challenge as well. 
 The key challenge here is to not focus on the extremes of completely-coupled and decoupled designs, but instead {\em explore
 the continuum of quasi-coupled designs in between to strike a balance between performance, feasibility and adaptability}.

\subsubsection{Choice of Connectivity Technology}
While the connectivity modality for RAN is given (LTE), we have a choice in the modality for backhaul connectivity.
The various options available to us include, sub-6 GHz technologies like WiFi and LTE, as well as high frequency technologies like mmWave (28 GHz, 60 GHz, 70-80 GHz, etc.) and FSO (free space optics).  

\begin{figure}
    \centering
    \includegraphics[width=0.8\columnwidth]{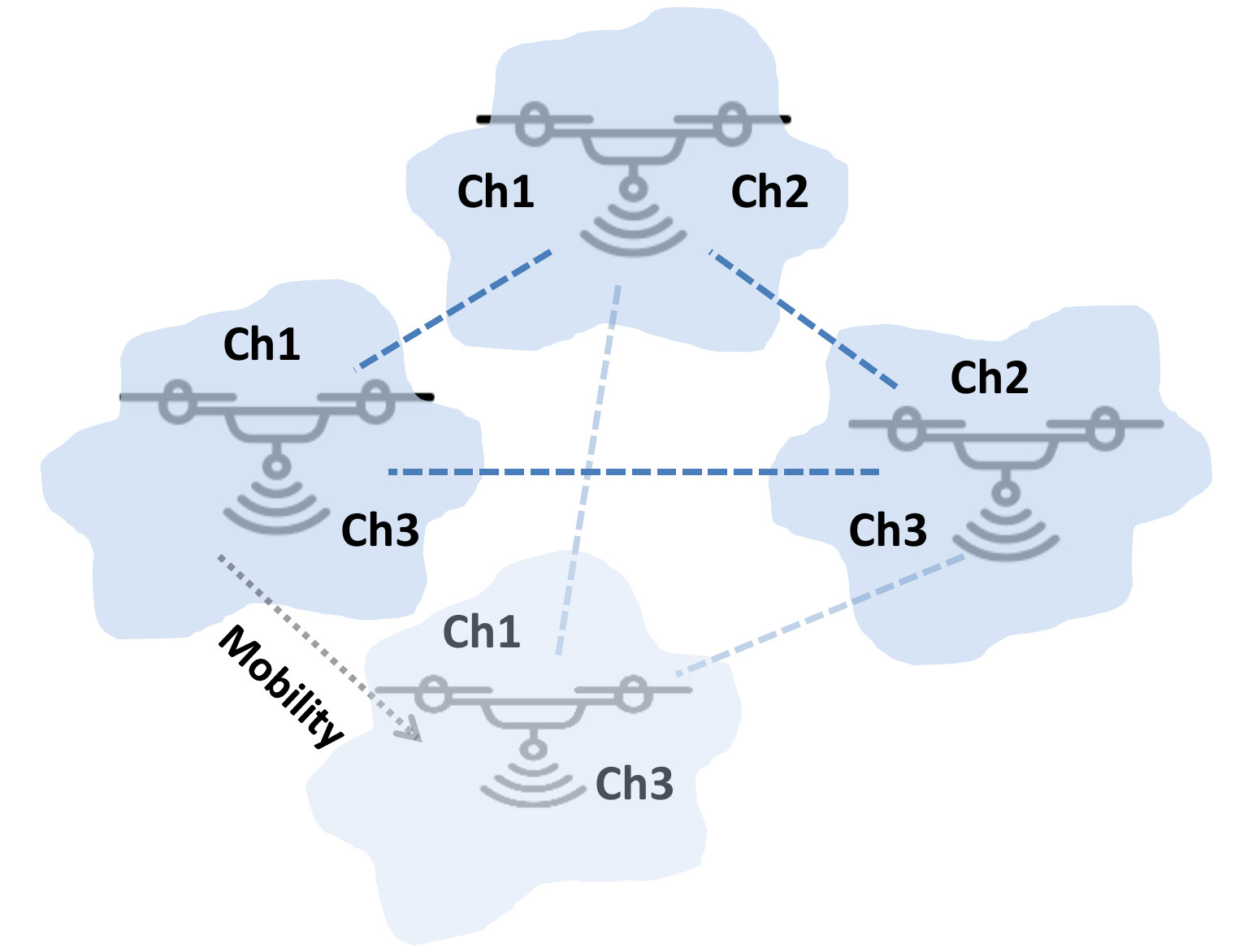}
    \caption{Omni-directional backhaul connectivity: Omni-directionality alleviates disruptions during mobility but requires intelligent spectrum allocation to avoid interference (each UAV establishes 2 parallel links on 2 channels using 2 radios).\label{fig:omnihaul}}
\end{figure}

{\bf mmWave/FSO vs. sub-6 GHz:} The biggest difference between these two categories lies in their need and ability to leverage directional wireless transmissions as well as the bandwidths offered (see Figs.~\ref{fig:omnihaul},\ref{fig:dirhaul}). Since sub-6 GHz bands have lower 
path loss (signal attenuation) compared to their high frequency counterparts, they can operate with omni-directional antennas and obtain
reasonable connectivity ranges. In contrast, mmWave relies on directional or phased array antennas (FSO employs lasers and photo-detectors) to form highly directional beams that increase signal energy in desired directions to compensate the increased attenuation in higher frequencies. On the positive side, the antenna form factor to realize such directional beams for mmWave is much smaller compared to that in lower frequencies. Also, larger contiguous bandwidth chunks (order of GHz) are available in higher frequencies compared to tens-hundreds of MHz in lower frequencies. Hence,  for high-altitude UAV networks, where supporting longer ranges between UAVs and larger coverage areas (higher traffic demand) is imperative, mmWave and FSO are more likely to serve the purpose better. With FSO offering a much higher bandwidth than mmWave, the former has been a primary focus in both Project Loon~\cite{Loon} and Aquila~\cite{Aquila}. 
However, for lower-altitude UAV networks with moderate range and capacity requirements, one could rely on the more economical omni-directional 
sub-6GHz technologies like WiFi and LTE to provide our backhaul connectivity modality. 
While radio equipment for mmWave (28 GHz, 70-80 GHz) and FSO are generally expensive and energy-draining to deploy on our low-altitude UAVs, an exception is the recent introduction of low-cost, mass-produced, unlicensed 60 GHz in WiFi (801.11ad) chipsets. 
Though 60 GHz faces increased oxygen absorption (20 dB/Km) compared to other mmWave bands, it offers higher bandwidth (1-2 Ghz) that could still be delivered over reasonable ranges (1-2 Km\footnote{Oxygen absorption loss of 20 dB/Km is compensated by an antenna array gain of 30-36 dB for 32-64 element arrays on either sides of the link.}) that may be sufficient for certain low-altitude UAV networks and forms a viable connectivity option.

{\bf LTE vs. WiFi:} LTE and WiFi are fundamentally different access/connectivity technologies. While LTE's synchronous transmissions (between transmitter and receiver) provide high spectral efficiencies, they are designed for a master-slave paradigm, where the base station controls all transmissions between itself and its users. While this is ideal for a single hop network,  it is challenging to deploy such a synchronous technology on a multi-hop network of UAVs, where each UAV needs to double up as a LTE relay (both as eNB and as UE) - the latter not being a mature technology yet. 
In contrast, WiFi is designed for unlicensed spectrum access and hence adopts an asynchronous transmission paradigm between nodes that makes it conducive for deployment on peer-peer UAV nodes. At the same time, being distributed in nature, WiFi incurs reduced performance efficiency. 
Further, WiFi operates in unlicensed spectrum and hence cannot guarantee quality-of-service owing to interference, unlike LTE that operators have optimized for licensed spectrum. Note that LTE can also aggregate unlicensed spectrum bands; e.g. CBRS 3.5 GHz bands, where available using LAA-LTE (License-assisted access LTE~\cite{ericsson,verizon,qualcomm}). Service reliability and guarantee could be a primary concern for operators when deploying a LTE network that caters to first responders in emergency situations (e.g. FirstNet~\cite{Firstnet}).

%

\subsubsection{Orientation and Provisioning}
\label{sec:config}
Based on the above discussion, the three viable connectivity options for backhaul include WiFi and LTE in sub-6 GHz, as well as WiFi in 60 Ghz. Each of these face their own set of challenges when it comes to orientation  as well resource provisioning at the UAV. 

\begin{figure}
    \centering
    \includegraphics[width=0.8\columnwidth]{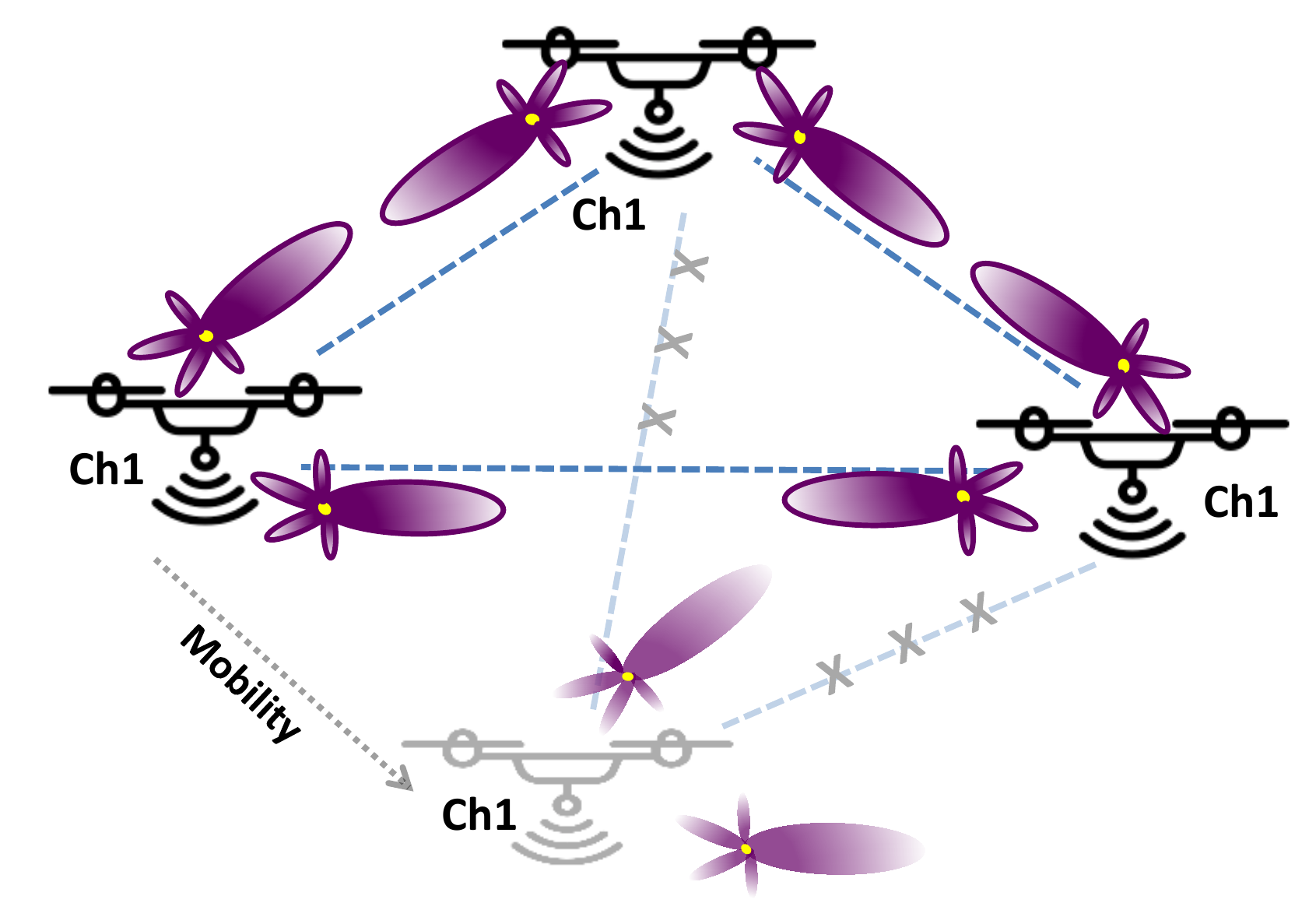}
    \caption{Directional backhaul connectivity: Directionality increases disruptions during mobility but also alleviates interference and hence need for additional spectrum (each UAV establishes 2 links on a single channel using 2 radios).\label{fig:dirhaul}}
\end{figure}

 {\bf Orientation:} In addition
to the UAV's position, its orientation  is equally important in determining backhaul connectivity, which in turn is coupled with its RAN connectivity. 
There are two components to UAV orientation: spatial orientation through beam selection (applies to directional transmissions) and physical UAV orientation (yaw/tilt). 
 While 60 GHz offers  higher bandwidth and leverages high directionality (beams with widths of a few degrees) to yield longer ranges, it comes at the expense of a fragile backhaul topology that is prone to link breakages and disconnections, even with minor movement of the UAVs (Fig.~\ref{fig:dirhaul}) -- the latter being an all too common characteristic in our environment.
Hence, discovering neighbors and finding the right configuration of the beams to use at the transmitter and receiver to even establish a link are critical challenges, not to mention the maintenance of the topology in the presence of UAV mobility, wind drifts, etc.
In contrast, the sub-6 GHz technologies do not incur these directionality challenges. 
However, irrespective of the technology employed, the physical orientation (yaw/tilt) of the UAV directly impacts both its backhaul and RAN connectivity. As the orientation of
the UAV changes, it changes the physical orientation of its antennas that are mounted (separately) for RAN and backhaul connectivity. A centralized controller might be necessary to establish and maintain efficient, well-connected backhaul topologies (with sufficient path diversity), especially when directionality is involved.

{\bf Provisioning:} With transmissions being highly directional in 60 GHz, the interference generated by a link on its neighbors is also negligible, allowing multiple links to operate in parallel on the same spectrum band. 
In contrast, the wireless interference generated by the omni-directional transmissions in sub-6 GHz does need to be handled through spectrum or other interference management approaches that comes at the expense of additional bandwidth (Fig.~\ref{fig:omnihaul}). 
With neighboring links interfering with each other, they have to be assigned on orthogonal spectrum (channels), thereby delivering lesser bits per unit spectrum. Further, with a single UAV communicating with multiple other neighboring UAVs to form a connected mesh network, care must be taken to ensure the finite spectrum available is appropriately allocated and reused across multiple hops. This is critical to avoid interference and maximize the amount of traffic that can be carried over the backhaul network.

We will jointly refer to the position, orientation and spectrum provisioning of the UAV as its configuration.



\subsubsection{Routing}
In addition to configuring the UAVs, one needs to intelligently route the traffic flows over the mesh backhaul so as to maximize the
amount of traffic demand that can be supported by the UAV network. However, these two aspects are not independent but rather tightly coupled.
The configuration of the UAVs determines the set of links as well as their capacities on the wireless mesh backhaul. 
Further,  the energy resources at the UAVs, in addition to their configuration information, also plays an important role in the longevity of the links.
Hence, by jointly optimizing the links along with the traffic routed over them, one can increase the longevity of connected backhaul topologies (minimize topology disruptions from UAVs going down for energy replenishment), thereby maximizing the aggregate flow routed over this backhaul.
However, realizing this joint optimization is a challenging problem and requires the collection of backhaul link configuration information at a central location to compute an efficient configuration and routing solution. Further, care must be taken to ensure that the solution can be executed in real-time to achieve their benefits in an inherently mobile backhaul network. 
While centralized routing solutions are more efficient than their distributed counterparts, the latter are more resilient to topology changes and disruptions from UAV dynamics.

Note that our discussions with respect to 60 GHz will also apply to other mmWave directional technologies (in 28 GHz, 38 GHz, 70-80 GHz, etc.), if they become viable in the future for low-altitude UAV networks. 
We summarize the essence of our backhaul connectivity discussions in Table~\ref{table:compbackhaul}.

\begin{table*}[t]
\centering
\caption{Comparison between EPC designs with varied EPC-RAN connectivity.}
\resizebox{\textwidth}{!}{

\begin{tabular}{l|c|c|c|c}
Challenge & \textsc{Legacy-Wired} & \textsc{Legacy-Wireless} & \textsc{Edge-OnDrone}  \\
\hline
\textbf{RAN-EPC link reliability} & High & Low & High  \\
\textbf{Reachability} & Poor & Fair & Good \\
\textbf{Capacity/Latency} & Fair & Poor & Good \\
\textbf{Scalability} & Poor & Fair & Good \\
\textbf{Overhead on UAVs} & Low & Low & High \\
\textbf{Network Induced Mobility} & Fair & Fair & Poor \\
\textbf{Seamless EPC-to-EPC functionality} & Poor & Poor & Poor \\
\textbf{Locating/Paging UEs} & Good & Good & Poor

\end{tabular}
}
\label{table:compepc}
\end{table*}

\begin{figure}
    \centering
    \includegraphics[width=\columnwidth]{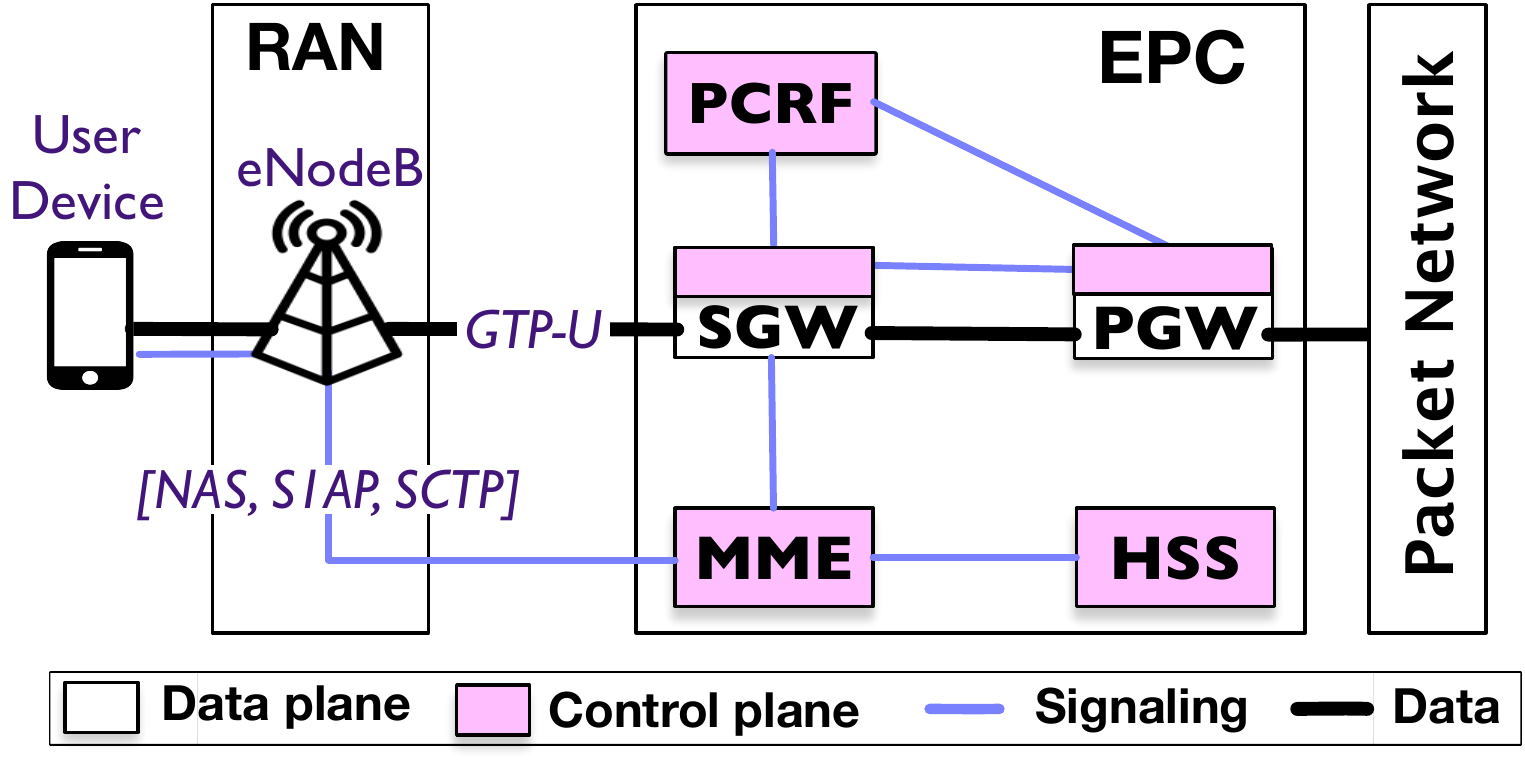}
    \caption{Legacy EPC Architecture.\label{fig:epc}}
\end{figure}

\subsection{Core Network Challenges}
So far we have discussed the challenges in establishing physical wireless connectivity on both the RAN and backhaul in our UAV networks.
However, for providing LTE connectivity, an equally important component is the logical connectivity provided by LTE's core (EPC) that serves as a middle-man between the base stations and/or the Internet. The EPC is responsible for managing various back-end functions (authentication, billing, management, mobility, etc.) required to maintain a smooth functioning of the LTE service as a whole. To foster better understanding, we first provide a short primer on EPC's key functionality, followed by the challenges faced in realizing an efficient EPC for our UAV network.

\subsubsection{EPC Primer}
Fig.~\ref{fig:epc} shows the network architecture of EPC, which is a distributed
system of different nodes or network functions (NFs)
that are required to manage the LTE network. The EPC consists
of data and control data planes: the data plane enforces
operator policies (e.g., DPI, QoS classes, accounting) on data
traffic to/from user devices, while the control plane provides
key control and management functions such as access control,
mobility and security management. eNodeBs (RANs)
are grouped into logical serving areas and connected to serving
gateways (SGW). The SGW is connected to an external
packet network (e.g. the internet) via a packet data network
gateway (PGW). PGW enforces most of data plane policies
(e.g., NAT, DPI) and may connect the core to other IP network
services (e.g., video server). The mobility management
entity (MME) is responsible for access control, security and
mobility functions (e.g., attach/detach, paging/handover) in
conjunction with the home subscriber server (HSS) database.

We will first discuss how to apply the existing EPC architecture as is to our UAV network and the drawbacks associated with such an approach.
 Then, we will discuss an alternate EPC architecture and contrast its pros and cons as well. 
 
\subsubsection{Legacy EPC Architecture}
\begin{figure}
    \centering
    \includegraphics[width=\columnwidth]{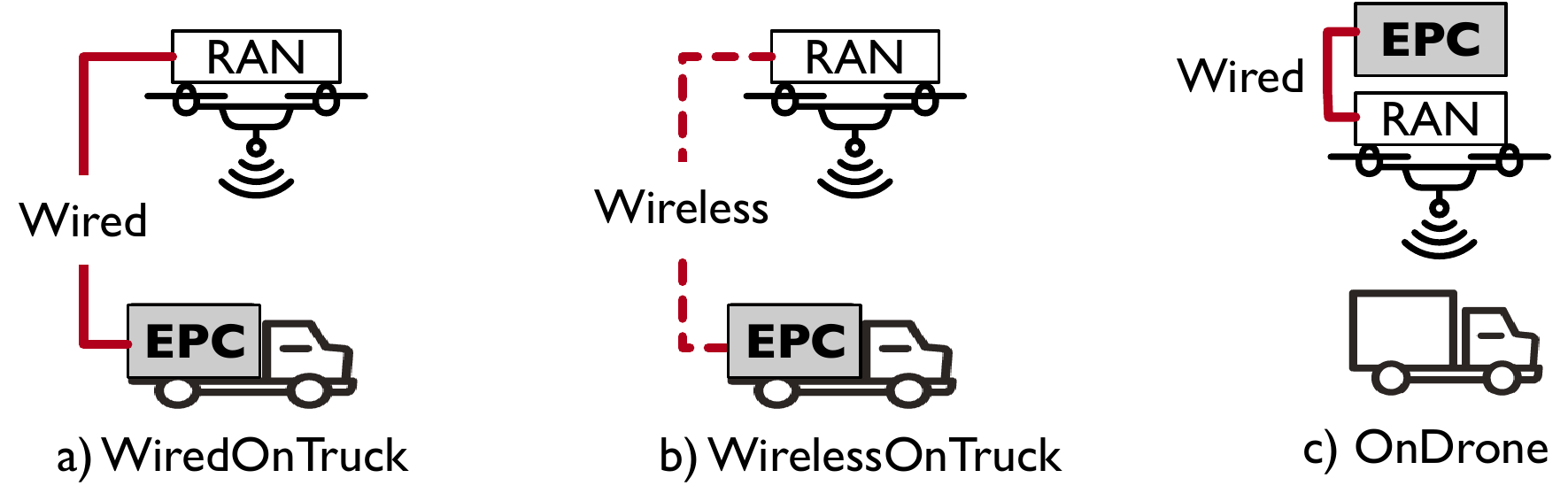}
    \caption{EPC Variants for UAV Networks.\label{fig:epcvariants}}
\end{figure}

Conventionally,  the EPC is a single wired network of distributed gateways deployed by the telecom operator to manage all the deployed base stations. 
When a UE sends/receives traffic, the EPC sets up a data session (bearer) between the eNB (to which the UE is connected) and S- and P-GWs. The PGW is responsible for interfacing with the public Internet as well as for routing traffic between different UEs in the same operator's network. 
The straight-forward way to apply EPC to our UAV network would be to collapse all the EPC network functions into a single node (EPC-in-a-box) and deploy this EPC node either on one of the UAVs or on the ground. EPC, being the orchestrator of the entire LTE RAN, deploying it on an inherently unreliable mobile platform like the UAV might be too risky. Hence, operators like AT\&T and Verizon, have chosen to deploy EPC on the ground in their current trials. There are two versions to deploying EPC on the ground: the connection between the EPC node on the ground and the UAVs in the air is either wireless (Verizon, Fig.~\ref{fig:vzwcow}) or wired (AT\&T, Fig.~\ref{fig:attcow1}) as shown in Fig.~\ref{fig:epcvariants}(a),(b).  

{\bf Reachability: }
In today's LTE networks, the connectivity between EPC and eNBs (RAN) is a reliable, wired network provisioned with sufficient bandwidth for catering to the UE traffic demands in both downlink and uplink. However, in our UAV network, deploying the EPC node on the ground makes the connectivity between EPC and eNBs (UAVs) wireless, which is inherently  unreliable. The wireless channel between the EPC node on the ground and the UAV may be subject to wireless artifacts  such as shadowing (building, trees, obstacles, etc.), multi path fading, etc. that degrade signal quality and can potentially cause disconnections. Since reachability to EPC is essential for a UAV to enable communication to/from its UEs, guaranteeing reachability to a ground EPC node from multiple UAVs that are deployed across a large region becomes a significant challenge.
One might have to deploy multiple EPC nodes on the ground to allow for reachability to all UAVs and to build robustness into the system, thereby adding to both the cost as well as reliance on ground deployments (limited flexibility).  

{\bf Choice of Connectivity Technology:} Similar to the backhaul challenge, we also need to determine an appropriate wireless technology for connecting the EPC node to the UAVs. To allow for reachability from a small set of ground EPC nodes to all UAVs in the air, one might need
to employ higher frequency (mmWave) technology that can provide the necessary beamforming/directionalty gain to deliver longer communication ranges. However, the latter would also be accompanied by the challenge of constantly tracking the direction of the beam with respect to each of the UAVs to maintain connectivity as they move. One could also employ FSO as the connectivity modality as is envisioned in Project Aquila~\cite{Aquila} for high-altitude UAV networks. However, this might be an over-kill for a low altitude UAV network given the associated cost.    

{\bf Capacity and Latency Bottleneck:} 
The EPC node on the ground becomes the routing focal point that ferries traffic not only between the UEs and the Internet but also between UEs within the UAV network.  Hence, even if the UAV backhaul is well-provisioned, have a small set of ground EPC nodes concentrates all traffic on the bachkaul towards these ground nodes, which in turn become the bottleneck. This would significantly degrade the capacity of the network as a whole. This is analogous to the gateway provisioning problem in wireless mesh networks~\cite{Multi-gateway}, where to avoid capacity bottlenecks, one would need to deploy multiple gateways that serve as entry/exit points into the wireless mesh backhaul. When bulk of the UAV network traffic is to/from
the Internet, which has to be accessed from the ground PoPs (points-of-presence), such a bottleneck is un-avoidable and needs to be addressed with the help of multiple PoPs and EPCs on the ground. This is the model followed by Project Loon~\cite{Loon} as it  aims to connect vast regions of previously-unconnected terrain to the Internet. However, for a low altitude UAV network deployed to provide on-demand connectivity to a smaller region in emergency situations, bulk of the traffic might involve communication and coordination between first responders and people affected in those local regions. In such scenarios, incurring the wireless capacity bottleneck to the EPC on the ground is un-warranted. 
In addition, UAV and UE mobility are highly pronounced in these networks, which leads to increased control signaling and associated latency between the ground EPC node and the UAVs. 
  
\subsubsection{Edge EPC Architecture}
 To counteract the challenges in deploying a legacy EPC architecture, we also consider an alternate, edge EPC architecture.
 Here, the intention is to push the  EPC functionality to the extreme edge of the core network, namely at the eNBs (UAVs) as shown in Fig.~\ref{fig:epcvariants}(c). In other words,
 the EPC is collapsed and located as a single, self-contained entity on each of the UAVs. While such an architecture completely eliminates all the drawbacks faced by the previous architecture resulting from wireless connectivity between EPC and eNBs, it faces a different set of challenges.  
  
{\bf Resource-challenged:}
An EPC consists of multiple network functions along with the interfaces and tunneling protocols between them. Further, most of these are state-ful network functions and consist of both control and data plane functionality.  These network functions, which used to be deployed by operators on specialized hardware, are now slowly migrating to a virtualization environment with the recent advances in NFV (network function virtualization)~\cite{vepc_ietf,epcopenflow,Scale}. Nevertheless, the compute resources consumed by these network functions is appreciable and becomes a concern when all the
EPC functionality is collapsed onto a single node. While deploying such an EPC node on the ground gives one access to both compute and energy sources for longer operational duration, this becomes a critical challenge when deployed directly on a UAV platform, which is highly
resource challenged to begin with. 
Deploying an EPC node on the UAV could significantly affect both its operational lifetime as well as the processing (control and data plane) latency of its traffic, thereby resulting in a highly reduced traffic capacity.   

{\bf Network Dynamics:}
Conventional EPC has a hierarchical structure, where a single PGW spans multiple SGWs, and a single SGW spans multiple eNBs. 
As the UE (in active mode) moves from one cell to another (handoff),
this is handled locally by its SGW. When a UE in idle mode (not actively communicating with the eNB)  moves across cells, it becomes harder to locate the UE. To address this, every UE has a tracking area (set of eNBs)  associated with it, which the EPC will use to page (all eNBs in tracking area) to locate it. When the UE moves out of its current tracking area, it notifies the EPC of its updated tracking area.   
Thus, UE mobility is handled seamlessly, which is one of the main features provided by the EPC. 

Network dynamics in the form of UE and/or UAV mobility forms a significant part of our operating environment. Hence, handling network (UE and UAV) mobility becomes all the more important in our case. However, by pushing EPC to the real edge (eNB) of the core network, providing seamlessly mobility now becomes substantially difficult. With the collapse of the hierarchical EPC architecture, one needs to now enable communication between the EPC entities on individual UAVs to enable seamless handoff across UAVs. In today's mobile
networks, a UE hardly moves across different PGWs within the same operator's network (a single PGW spans a significantly large area - hundreds of miles). When such an event does happen, the connection is terminated with the existing PGW and re-established with the new PGW causing service disruption.   However, with EPCs located on each UAV  in our environment, such events are the norm rather than an exception. Hence, it becomes critical to enable seamless EPC-EPC communication for handling mobility in the edge EPC architecture. This is needed to also handle UAV mobility, i.e.  when one UAV goes down for re-charge and is replaced by another UAV -- a migration of state from one UAV (EPC) to the other is imperative. 

{\bf Locating UEs:}
In legacy EPC architecture, the PGW serves as the central ingress and egress points for traffic to all UEs and keeps track of the various bearers (sessions) to each UE. Along with the ability to page idle UEs over large tracking areas, it is fairly straight-forward to locate any UE in the network. This is however, a challenge for our edge EPC architecture, where there is no single PGW that spans all the UAVs (eNBs).  
Further, handling UE mobility in idle mode is even more challenging. Since 
the notion of tracking area disappears (due to collapsed EPC), locating a UE when in idle mode appears to be infeasible, prompting the need for new or adapted mobility mechanisms.

We summarize the pros and cons of these two EPC architectures as they apply to our UAV networks in Table~\ref{table:compepc}.

\begin{figure}
    \centering
    \includegraphics[width=\columnwidth]{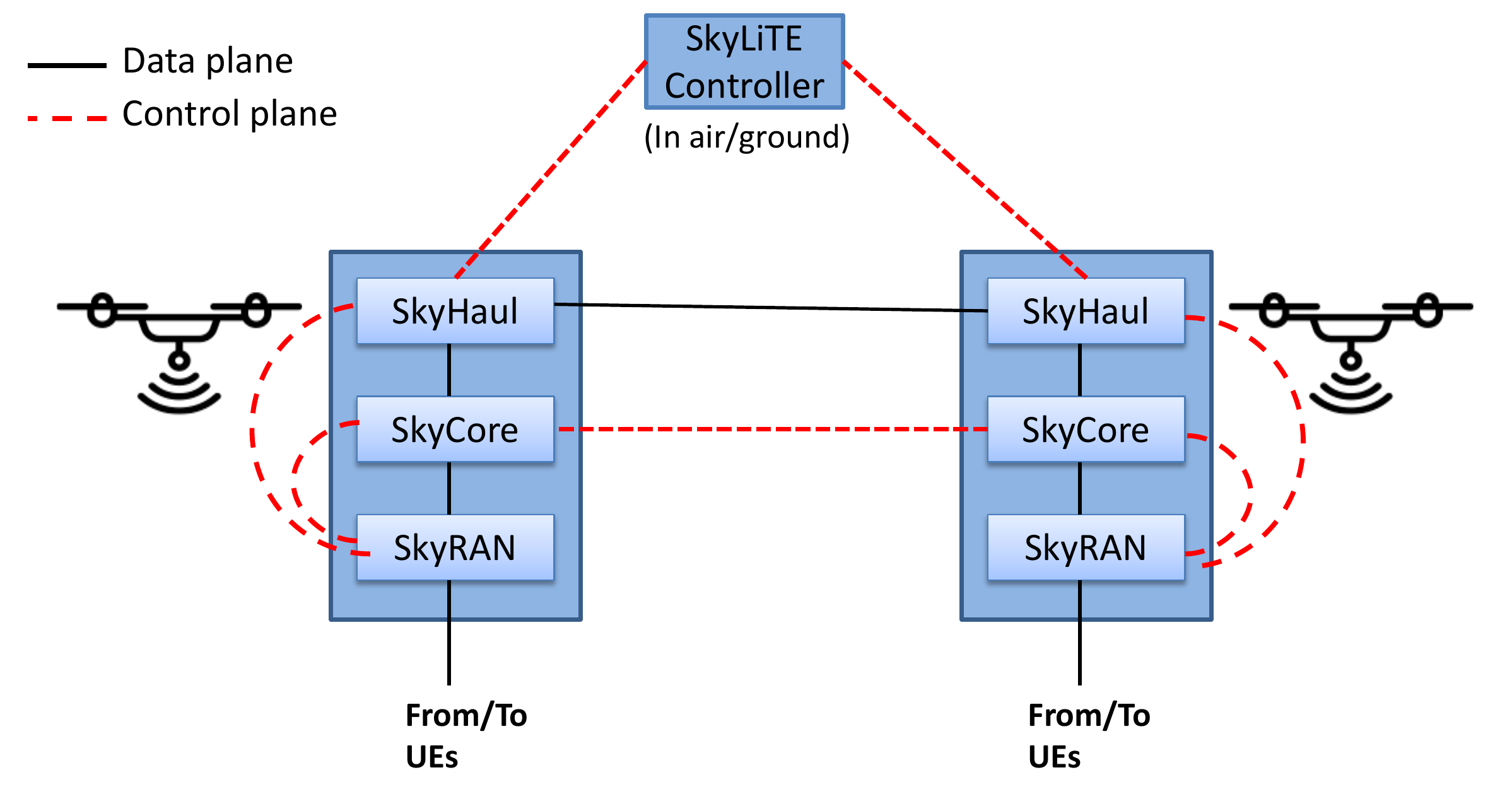}
    \caption{SkyLiTE Overview.\label{fig:skylite}}
\end{figure}


\section{{\large{\bf \system}}: An End-End Design}
\label{SKY}


The previous section highlighted the numerous, inter-twined challenges that spanned the layers of RAN, backhaul and core network, in
the design of an end-end UAV network for providing LTE connectivity. The optimal design choice for the UAV platform, radio interfaces, access/core techniques and optimizations, etc. would vary from one set of application requirements to another (e.g. low vs. high altitude, small vs. large coverage, short vs/ long timespan),
 and there is no single design that can cater effectively to all applications. Nevertheless, to give the readers a good starting point, we articulate the high-level overview of an end-end design called \system that explores some of the key functionalities. \system is composed of three essential modules as shown in Fig.~\ref{fig:skylite}:
 \sysran, \syshaul, and \syscore that are responsible for handling the inter-twined challenges in the RAN, backhaul and core network layers respectively. 
 We will present the key functionality of each of these modules as well as the interactions between them that together help realize our \system system. Specific solutions and algorithms underpinning each of these modules will be covered in separate articles.

\subsection{Overview}
 Recall that our application requirement is to provide optimized LTE coverage for a designated/given terrestrial area, given a certain number of drones. 
 \system's high level approach consists of two components: a bootstrapping phase and a periodic update phase. 
 In the bootstrapping phase,  \system partitions the given area into smaller coverage zones, where each of them will be covered by a single UAV.
 It then determines the configuration of each of the UAVs to optimize both RAN as well as backhaul connectivity and routing jointly. Here, the design targets the support of a minimum desired traffic demand from every UAV to every other UAV in the backhaul during the bootstrapping phase.  Also, additional UAVs may be minimally deployed to provide a high-capacity, reliable, mesh backhaul in the process. 
 After the bootstrapping phase, once the location of the UEs and traffic demands between different entities in the UAV network as well as 
 to/from the Internet 
 are estimated, every subsequent update phase will reconfigure the UAVs to cater to the spatio-temporally observed network  and traffic conditions. Hence, based on changing conditions, UAVs can also be removed from over-provisioned areas as well as added to under-provisioned areas as needed. 

\subsection{Hybrid Design}
 To determine the UAV configuration as well as routing on the backhaul, we need to revisit our discussion/tradeoff on joint RAN and backhaul design -- it is ideal to realize a completely joint design but impractical to execute it in practice.
 \system addresses this tradeoff by exploring in two complementary directions - hardware (a flexible UAV platform) and software (pseudo joint optimization) approaches.
  
 {\bf Hardware:} Note that the connectivity on the UAV is achieved throguh two sets of antennas, one for RAN and other for backhaul.
 If the UAV platform can support decoupled antenna mounts for RAN and backhaul connectivity, this would weaken the strong coupling between RAN and backhaul optimization -- orientation of the RAN link will no longer affect that of the backhaul link and vice versa.
 Hence, the UAV can have separate RAN and backhaul configurations. The position of the drone will still be common to these two configurations. However,  a sequential optimization may be sufficient to handle this coupling, which is less stringent than before. 
 First, the UAVs will independently (locally) configure
 themselves for RAN optimization; then given their RAN configuration as well as their RAN traffic demands, their backhaul
 configurations (position fixed from RAN configuration) as well as traffic routing over the backhaul are determined. If the backhaul optimization
 requires additional backhaul-specific UAVs to be deployed, these UAVs will also have their position determined as part of their configuration.

 {\bf Software:} Often the antenna mounts for both RAN and backhaul are coupled to the UAV's frame and hence dependent on one another. In this case,
 a more closely coupled optimization is needed between the RAN and backhaul. As before each of the UAVs will first determine their RAN configuration. However, instead of picking the most optimal RAN configuration, each UAV will determine an acceptable set (based on some RAN performance requirements) of RAN configurations.Then given these feasible set of RAN configurations for each UAV, the backhaul configuration for each UAV is jointly selected  from their respective feasible set to optimize backhaul performance.  Having a larger configuration space for each UAV after RAN optimization allows \system to strike a balance between RAN and backhaul performance without being biased towards the former. 

Thus, \system employs a hybrid design, where it adopts a completely decentralized RAN optimization that is executed locally at each of the UAVs in parallel, while it adopts a centralized backhaul optimization for determining the configuration of the UAVs and routing on the backhaul. 
A decentralized approach to RAN optimization allows \system to track and optimize for UE dynamics that vary at fine time scales (seconds), while being highly scalable in a large UAV network. In contrast, with the backhaul dynamics varying at coarse time scales (minutes), \system leverages a centralized approach to realize optimal backhaul configuration and routing. Such a hybrid design allows \system to strike an effective balance between performance and scalability. 

\subsection{\bf \large \sysran} 
 This module runs in each UAV locally and is responsible for determining the UAV's configuration for optimized RAN performance in its designated terrestrial zone. 
 The terrestrial zone that needs to be covered by the UAV with some performance requirements (e.g. above a certain SNR/rate for each UE in the zone) and the configuration capabilities (movement/position, transmit power, antenna pattern (tilt/yaw)) of the UAV are known. 
 Given this, there exists a set of points in the 3D airspace (along with appropriate UAV transmit parameters for each point) from which the UAV will be able to deliver required coverage in the designated zone.
 \sysran first estimates this operational airspace for the UAV. Next, it figures out where  to specifically position itself in this airspace so as to deliver optimized coverage performance for the current set of UEs in its coverage zone. Thus,  while the operational airspace is constructed generic to the coverage zone, the eventual positioning of the UAV is optimized for the location of its UEs in the zone. 

 To accomplish the second step, \sysran leverages the LTE RAN and its synchronous transmission characteristics to automatically localize its devices without relying on their GPS functionality. It does so by sampling a few locations in its operational airspace and uses the LTE's reference signal channel measurements from the UEs to estimate the range (from time of flight) to the UEs from each of those points; then knowing the UAVs own location (using GPS) at the different points, employs trilateration to solve for the location of the UEs. 

 Once the location of the UEs is estimated, \sysran then constructs an RF map of its operational airspace for each of the UEs. In essence,
 it creates a map that predicts the RF signal strength at the UE with high accuracy for each of UAV's position in the operational airspace. To create this map as quickly as possible (without exhaustively conducting measurements to UEs from all points in the airspace), \sysran leverages the location of all its UEs to design a hierarchical
 measurement trajectory for the UAV, whereby the UAV first samples different points in the airspace at a coarse level; then based on the statistics of its coarse sampling, it employs a fine sampling of specific regions as needed to construct an accurate RF map for all its UEs.  
 Note that, with each UAV running its \sysran module in parallel, it is possible that the operational airspace of multiple UAVs overlap. In such cases, the UAVs follow an implicit priority ordering to avoid collision. Here, when two UAVs are within a minimum separation range (MSR) of each other, the one with the lower priority will stop in its trajectory and wait for the higher priority UAV to move outside the MSR, before carrying on with its trajectory. 

 Finally, with the RF maps for all the UEs estimated, \sysran solves an optimization problem (based on desired coverage objective) to either determine the optimal configuration of operation (when RAN and backhaul are weakly coupled) or narrow down a set of efficient configurations (when RAN and backhaul are tightly coupled).

\subsection{\bf \large\syshaul}
The \syshaul module runs in each of the UAVs and is responsible for coordinating the optimization of backhaul connectivity. 
For its centralized backhaul optimization, \system can leverage the same control channel and associated controller that is used for the UAV network's C\&C. \system's controller adopts an SDN (software-defined networking) approach to gather all the relevant backhaul information (from the \syshaul module in its UAVs) necessary to run its centralized optimization, as well as deliver the resulting computed configurations and traffic routing policies back to its UAVs through their \syshaul, which is responsible for the execution of the routing policies. 

\syshaul periodically gathers information regarding the incoming (to its RAN) and outgoing (from its RAN) traffic demand at the UAV, backhaul capabilities (energy resources, antenna mount, number of radio interfaces, connectivity technology, etc.)  of the UAV, as well as the candidate UAV configurations based on RAN optimization. The \syshaul at each UAV then communicates this information to \system's controller, which then uses this information as input along with the remaining number of UAVs available for deployment to run its backhaul optimization. The goal of the optimization is to configure the backhaul to support the observed traffic demand from the RAN, while deploying a minimum number of  UAVs as needed.
Note that since the UAVs are locally optimized from a RAN coverage perspective, they may not be optimally connected to each other to form a high-capacity, reliable mesh backhaul. Hence, in such cases, the controller will automatically determine the need to add or prune UAVs as needed.
While the controller runs its optimization periodically, in the event of un-planned UAV dynamics (UAVs going down for energy replenishment), the controller will invoke its optimization for a backhaul reconfiguration on-demand. 

In the case of sub-6 GHz backhaul connectivity, the controller also determines the appropriate allocation of wireless channels at each of the UAVs so as to minimize/avoid wireless interference between neighboring transmitters and maximize the traffic flow that can be routed over the backhaul.   On the other hand, for high frequency directional technologies like mmWave and FSO, wireless interference is less of a concern. However, the the controller now determines the appropriate beam orientations for each of the UAVs so  as to create desired high-gain directional, backhaul links between UAVs. 
\syshaul prefers to leverage a cost-effective, high-bandwidth mmWave technology like 60 GHz for its backhaul, when moderate backhaul ranges of 1-2 Km are sufficient. 

\subsection{\bf \large \syscore}
With the challenges of reachability, capacity and latency in deploying legacy EPC on the ground and away from the UAV network, \syscore adopts
the edge EPC architecture as shown in Fig.~\ref{fig:skylite}. \syscore collapses the entire EPC and pushes it to the edge of our network, namely at each of the UAVs themselves, where the RAN also resides.

With every UAV now running its own EPC agent, even a simple eNB-eNB handoff across two UAVs now becomes a inter-MME (MME-MME) handoff, which needs to be accomplished across two different EPC agents. Hence, \syscore enables a new control/data interface to enable EPC-EPC signaling and communication directly between UAVs to handle mobility right at the edge. To reduce its compute footprint on the UAV, \syscore adopts a software refactoring approach to eliminate distributed EPC interfaces and collapse all distributed functionalities into a single logical entity. It realizes this by transforming the distributed data plane functions into
a series of switching flow tables and associated switching actions (corresponding to network functions like GTP encapsulation/decapsulation, charging, etc.). It also reduces control plane signaling and latency by pre-computing and storing (in-memory) several key attributes relating to security keys, QoS profile, etc. for the UEs that can be accessed locally and quickly in real-time without any computation.

With the EPC being located on each UAV, the tracking area for a UE corresponds to a single eNB (UAV) in our case (compared to a set of eNBs in legacy EPC). To address the challenging problem of locating the UE during mobility,  the \syscore agent at any UAV 
automatically broadcasts the detection of a new (incoming) UE (either in active or idle mode) to the other \syscore agents in the UAV network. While the detection of an active mode UE is simple, this is difficult for an idle mode UE.
However, since the tracking area of the UE changes as it moves from one UAV to another, though the UE is in idle mode, it will request for a tracking area update. Upon receiving the latter, the MME in the \syscore agent of the new UAV, will be able to detect this new UE's arrival. 
The HSS in each \syscore agent maintains the location (anchoring \syscore agent)  of all UEs in the network. Hence, when a \syscore agent sends a UE location update, the agents in other UAVs update their HSS accordingly. Thus, whenever traffic needs to be sent from a \syscore agent to a specific UE located at another UAV, the HSS will reveal the destination \syscore agent at which the UE is anchored and to whom the traffic has to be routed. The actual routing path to be taken by the traffic on the mesh backhaul is then determined by the \syshaul agent at the UAV. 

It must be noted that while \syshaul adopts a centralized approach to optimization, \syscore follows a distributed approach similar to \sysran.
The rationale behind such a design stems from the fact that both \sysran and \syscore deal directly with individual UEs, whose time scale of dynamics warrants a local, distributed approach. Im contrast, \syshaul that deals with aggregate traffic (from multiple UEs), whose coarse time scale of dynamics allows for a centralized approach.

\section{Prototyping \large{\bf \system}}
\label{PROTO}

\begin{figure}
    \centering
    \includegraphics[width=\columnwidth]{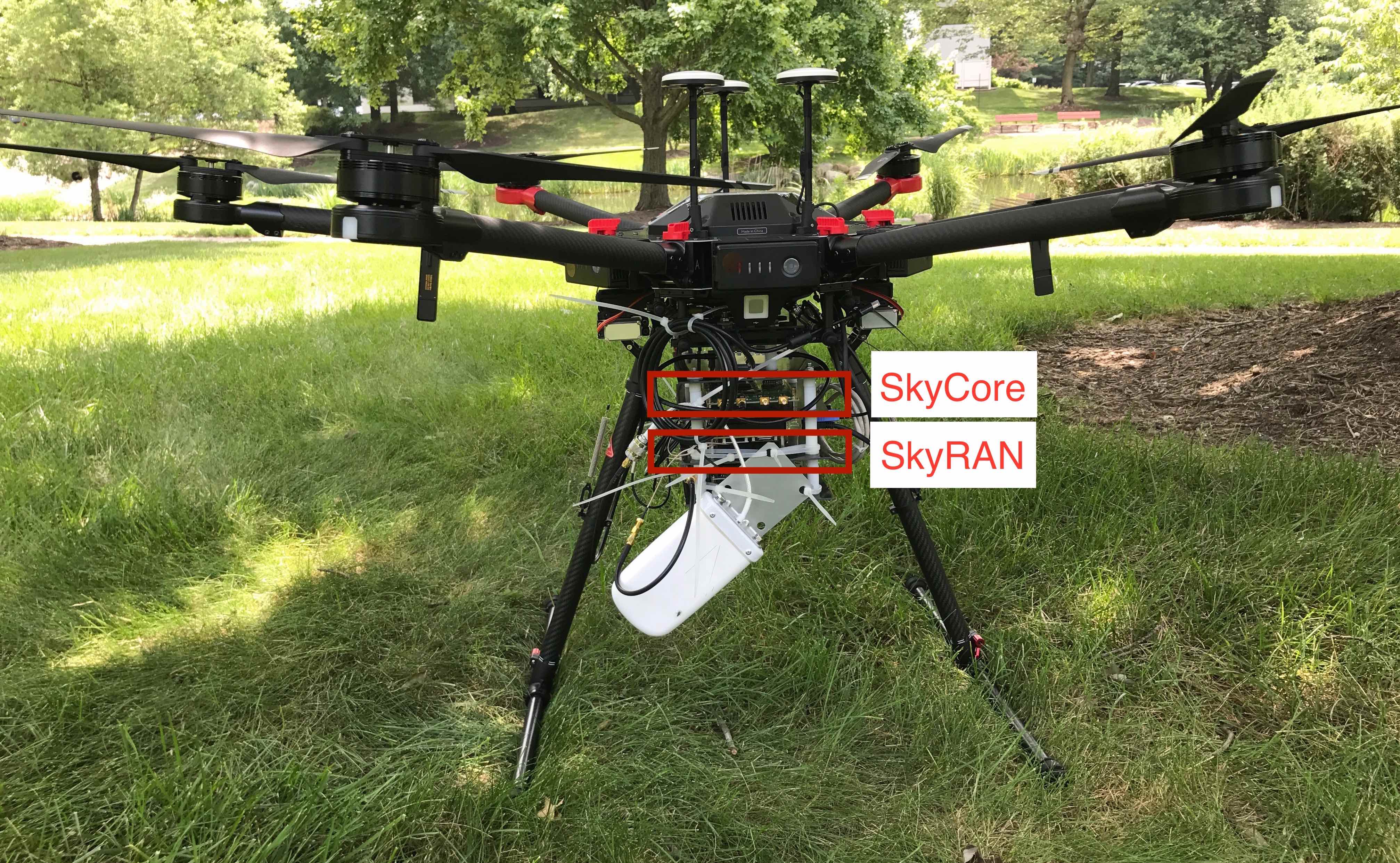}
    \caption{\system Prototype.\label{fig:prototype}}
\end{figure}

We have built an initial version of \system with the \sysran and \syscore modules and successfully deployed it on a DJI hexacopter (as shown in Fig.~\ref{fig:prototype}) that provides LTE connectivity to smartphones on the ground. 
The \sysran and \syscore modules are built upon software-defined versions of RAN and Core network stacks like Open Air Interface~\cite{oai} and OpenEPC~\cite{openepc} with appropriate modifications needed to realize the desired features, while maintaining standards compliance. 
This demonstrates that with the right optimizations, it is possible to deploy and operate a self-contained mobile network on the UAV directly.
We are in the process of testing and evaluating the performance of the individual modules in conjunction with that of the UAV platform as well as adding the various proposed optimizations for each module. The results of this study will be disseminated in subsequent articles.
We will then extend \system to accommodate multiple UAVs with the addition of the \syshaul module.

\section{Discussions}
\label{DISC}
\subsection{Additional Considerations for Fixed-Wing Aircrafts}
\label{FIXED}
The considerations for fixed wing aircrafts include those highlighted in \system for rotary wing aircrafts. Hence, \system's design elements can be leveraged in LAP networks that employ the fixed wing aircrafts as UAVs. These UAVs offer better endurance and payload capacity compared to their rotary wing counterparts. However, they are subject to constant mobility, which imposes additional
challenges and warrants special design considerations. While mobility of the UAVs cannot be avoided, it can be controlled and hence offers
another degree of freedom in the optimization of the LAP network through path planning of the UAVs. This creates a more involved joint 
optimization  of RAN and backhaul connectivity -- compared to just the UAV configuration in rotary winged aircrafts, fixed wing aircrafts have to plan and control the 
trajectory of multiple UAVs simultaneously, the dynamics of which, significantly impacts both RAN and backhaul connectivity.

\subsection{Applicability to HAP Networks}
\label{HAP}
Where there are several similarities in network design between LAP and HAP networks, there are also significant differences
that calls for a different perspective in weighing challenges and design elements appropriately. We mention some of the key differences here as well as the applicability of \system's design elements. 

{\bf RAN:} In a LAP network, obstacles and multi path scatterers between the UE and UAV have a substantial impact on the LTE link performance. In contrast, in a HAP network, the large path loss (due to distance) between UAV and the UE overshadows the impact of shadowing and fading when determining LTE link performance. Consequently, fine-grained RAN coverage optimization based on the UEs is less relevant in a HAP network. However, \sysran's mechanisms related to localizing the UEs as well as coordinating multiple-UAV for a coarse coverage optimization are still useful for HAP networks.

{\bf Backhaul:} 
From an optimization stand-point, this forms the most important part of the UAV network that needs to be intelligently and dynamically provisioned as well as maintained. In addition to the applicability of \syshaul's design components, HAP networks also face another challenge in their backhaul design, namely the constant mobility of the UAVs themselves, similar to that faced in fixed wing aircrafts. Recall that the UAVs in these networks leverage the stratified air currents in the atmosphere to travel and hence are prone to constant movement. Hence, the backhaul must be capable of constantly predicting the position of its UAVs to optimize the connectivity of its backhaul ahead of time. This adds another dimension (of time) to backhaul optimization compared to that already considered in \syshaul. Recently, \cite{Loon-SDN} articulated the notion of a spatio-temporal SDN for designing such a backhaul.

{\bf Core:}
Although today's high-altitude UAV networks envision to have their EPC on the ground, we believe the benefits of a \syscore design for edge EPC significantly outweigh its drawbacks, and is equally applicable to a high-altitude UAV network as well.

\section{Conclusion}
\label{CONCL}
UAV networks are ushering in a novel paradigm for wireless connectivity with a host of new applications and services. 
However, leveraging them to their full potential requires one to first understand the various challenges that underline their design.
This has been the prime focus of this work, which has tried to unravel the various inter-twined challenges that span across the layers of access, core network and backhaul in designing a low-altitude UAV network for providing LTE connectivity. We have also presented an end-to-end design called \system that can serve as a framework or starting point for the design and optimization of such UAV networks. Through this document, we also hope to engage the broader research community as well as industry towards addressing these challenges and making these UAV networks viable in practice.

\bibliographystyle{ieee}
\bibliography{paper,form}

\begin{thebibliography}{10}

\bibitem{Loon}
``Google x: Project loon,'' \url{https://x.company/loon/}.

\bibitem{Aquila}
``Facebook project aquila,''
  \url{https://code.facebook.com/posts/348442828901047/aquila-what-s-next-for-high-altitude-connectivity-/}.

\bibitem{AttCow}
``At\&t cell on wings,'' \url{http://about.att.com/innovationblog/cows_fly}.

\bibitem{VzwCow}
``Verizon cell on wings,''
  \url{https://newatlas.com/verizon-drones-internet-trials/45818/}.

\bibitem{AttCowPR}
``At\&t's lte-equipped flying cow in puerto rico,''
  {\url{https://www.fiercewireless.com/wireless/at-t-deploys-lte-equipped-flying-cow-drone-puerto-rico}}.

\bibitem{EU-Absolute}
S.~Chandrasekhara et. al.,
\newblock ``{Designing and Implementing Future Aerial Communication
  Networks},''
\newblock {\em {IEEE Communications Magazine}}, vol. 54, no. 5, pp. 26--34, May
  2016.

\bibitem{uav-link1}
N.~Ahmed, S.~S. Kanhere, and S.~Jha,
\newblock ``On the importance of link characterization for aerial wireless
  sensor networks,''
\newblock in {\em IEEE Wireless Communications Magazine}, May 2016.

\bibitem{uav-link2}
B.~V.-D. Bergh, A.~Chiumento, and S.~Pollin,
\newblock ``Lte in the sky: Trading off propagation benefits with interference
  costs for aerial nodes,''
\newblock in {\em IEEE Wireless Communications Magazine}, May 2016.

\bibitem{ericsson}
``Lte license assisted access,''
\newblock {\em Ericsson}, 2015.

\bibitem{verizon}
``Verizon to trial spidercloud lte-u scalable in-building system for
  enterprises and venues,'' \texttt{http://www.spidercloud.com/news/
  press-release/verizon-trial- spidercloud-lte-u-scalable-building-
  system-enterprises-and-venues}, 2016,
\newblock [Online; Accessed March 14 2016].

\bibitem{qualcomm}
``{LTE} in unlicensed spectrum: Harmonious coexistence with {Wi-Fi},''
\newblock {\em Qualcomm Research}, June 2014.

\bibitem{Firstnet}
``Firstnet: For public safety, by public safety,''
  \url{https://www.firstnet.com}.

\bibitem{Multi-gateway}
S.~Lakshmanan, K.~Sundaresan, and R.~Sivakumar,
\newblock ``{Multi-gateway Association in Wireless Mesh Networks},''
\newblock {\em {Ad-Hoc Networks}}, vol. 7, no. 3, May 2009.

\bibitem{vepc_ietf}
D.~King, Liebsch M., Willis P., and Ryoo J.,
\newblock ``Virtualisation of mobile core network use case,''
  \url{http://tinyurl.com/mvveqyp}.

\bibitem{epcopenflow}
J.~Kempf, B.~Johansson, S.~Pettersson, H.~Luning, and T.~Nilsson,
\newblock ``Moving the mobile evolved packet core to the cloud,''
\newblock in {\em Wireless and Mobile Computing, Networking and Communications
  (WiMob), 2012 IEEE 8th International Conference on}.

\bibitem{Scale}
A.~Banerjee, R.~Mahindra, K.~Sundaresan, S.~Rangarajan, and S.~Kasera,
\newblock ``Scaling the lte control plane for future mobile access,''
\newblock in {\em ACM CoNEXT}, Dec 2015.

\bibitem{oai}
``Open air interface: 5g software alliance for democraticing software
  alliance,'' \url{http://www.openairinterface.org}.

\bibitem{openepc}
``Openepc: The open epc project,'' \url{https://www.openepc.com}.

\bibitem{Loon-SDN}
B.~Barritt, T.~Kichkaylo, K.~Mandke, A.~Zalcman, and V.~Lin,
\newblock ``Operating a uav mesh \& internet backhaul network using
  temporospatial sdn,''
\newblock in {\em IEEE Aerospace Conference}, June 2017.

\end{thebibliography}
\normalsize
\end{document}